\documentclass[reprint,superscriptaddress,aps,prl,nobibnotes,longbibliography]{revtex4-2}
\usepackage[T1]{fontenc}
\usepackage{amsmath,amssymb,bm}
\usepackage{amsfonts,mathrsfs}
\usepackage{graphicx}
\usepackage{xcolor}
\usepackage{color}
\usepackage{babel}
\usepackage{textcomp}
\usepackage{dsfont}
\usepackage{amsthm}
\usepackage{ulem}
\usepackage[unicode=true,
            pdfusetitle,
            bookmarks=true,
            bookmarksnumbered=false,
            bookmarksopen=false,
            breaklinks=false,pdfborder={0 0 0},
            backref=false,
            colorlinks=true,
            linkcolor=blue,
            citecolor=blue,
            urlcolor=blue]{hyperref}

\makeatletter

\theoremstyle{plain}

\makeatother

\begin{document}

\title{Interlocked Time Crystal in Coupled Spin-1/2 Ensembles under Local Dissipation}
\author{Zhen-Huan Yang}
\affiliation{Key Laboratory of Atomic and Subatomic Structure and Quantum Control (Ministry of Education),  Guangdong Basic Research Center of Excellence for Structure and Fundamental Interactions of Matter, and  School of Physics, South China Normal University, Guangzhou 510006, China}

\author{Zhen-Tao Liang}
\affiliation{Key Laboratory of Atomic and Subatomic Structure and Quantum Control (Ministry of Education),  Guangdong Basic Research Center of Excellence for Structure and Fundamental Interactions of Matter, and  School of Physics, South China Normal University, Guangzhou 510006, China}
\affiliation{Guangdong Provincial Key Laboratory of Quantum Engineering and Quantum Materials,  Guangdong-Hong Kong Joint Laboratory of Quantum Matter, and Frontier Research Institute for Physics,\\  South China Normal University, Guangzhou 510006, China}
\affiliation{Quantum Science Center of Guangdong-Hong Kong-Macao Greater Bay Area, Shenzhen, China}

\author{Dan-Bo Zhang}
\email{dbzhang@m.scnu.edu.cn}
\affiliation{Key Laboratory of Atomic and Subatomic Structure and Quantum Control (Ministry of Education),  Guangdong Basic Research Center of Excellence for Structure and Fundamental Interactions of Matter, and  School of Physics, South China Normal University, Guangzhou 510006, China}
\affiliation{Guangdong Provincial Key Laboratory of Quantum Engineering and Quantum Materials,  Guangdong-Hong Kong Joint Laboratory of Quantum Matter, and Frontier Research Institute for Physics,\\  South China Normal University, Guangzhou 510006, China}

\date{\today}

\begin{abstract}
Multilevel dissipative systems can exploit multiple local transitions and coherence channels to generate nonstationary time-crystalline dynamics.
Here we show that an analogous mechanism can be synthesized without enlarging the local Hilbert space, by coupling two locally pumped and decaying spin-$1/2$ ensembles into a composite dissipative unit.
Neither ensemble supports an autonomous oscillatory phase; instead, opposite pump-decay imbalances and inter-ensemble exchange coupling can lead to a single interlocked time crystal with a fixed internal phase relation and no single-ensemble counterpart. The  time-crystalline character is consistently established through the mean-field analysis, exact calulation of Liouvillian spectra at finite size, and temporal correlations with cumulant expansion. Our work establishes a route to dissipative time-crystalline order in which coupling between simple two-level subsystems generates the effective internal structure otherwise provided by multilevel constituents. 
\end{abstract}

\maketitle

\textit{Introduction---}
Continuous time crystals are nonequilibrium phases that spontaneously break continuous time-translation symmetry, yielding persistent collective oscillations in the thermodynamic limit~\cite{Wilczek2012QuantumTC,Shapere2012ClassicalTC,Bruno2013NoGo,Watanabe2015Absence,Sacha2018Review,Hannaford2022Decade}.
In open many-body systems, dissipation can stabilize this order, producing dissipative and boundary time crystals~\cite{Nakatsugawa2017QuantumTimeCrystalDecoherence,Iemini2018BoundaryTC,Buca2019Dissipation,Carollo2022ExactBTC,Seibold2020PhotonicDimer,Alaeian2022ExactMultistability,Bakker2022TwoModeKerr,Yang2025EmergentCTC}.
Experiments span ultracold atom--cavity systems~\cite{Kongkhambut2022ContinuousTC,Dreon2022SelfOscillatingPump,Cosme2025TorusBifurcation}, thermal atomic gases~\cite{Wu2024RydbergDTC,Arumugam2025StarkModulatedRydbergDTC,Huang2025SpinGasCTC}, solid-state spins~\cite{Chen2023InherentTC,Greilich2024RobustCTC}, and photonic or polaritonic platforms~\cite{CarraroHaddad2024PolaritonCTC,Chen2026PhotorefractiveCTC}.

While most realizations are characterized by population, density, or field oscillations, another possible manifestation is transverse collective coherence.
Cavity-mediated atomic ensembles and steady-state superradiant lasers demonstrate macroscopic phase coherence and collective-dipole synchronization as robust collective phenomena~\cite{Bohnet2012SuperradiantLaser,Xu2015ConditionalRamsey,Zhu2015SynchronizationDipoles,Weiner2017SuperradiantSync,Roth2016ActiveAtomicClocks}.
More recently, many-body ensembles of finite-dimensional quantum limit-cycle oscillators can spontaneously develop transverse collective coherence, a phenomenon termed macroscopic quantum synchronization~\cite{Nadolny2023MacroscopicSync}.
At mean field, these synchronized states share with continuous time crystals the emergence of autonomous oscillations with a spontaneously selected phase, motivating us to examine how local Hilbert-space structure shapes collective coherent dynamics.

In a three-level oscillator, the spin-1 ladder operator contains two adjacent transitions, $|0\rangle\leftrightarrow |1\rangle$ and $|1\rangle\leftrightarrow |2\rangle$, so the collective transverse amplitude is built from two local coherences whose mean-field responses are shaped by transition population differences set by local gain and loss~\cite{Nadolny2023MacroscopicSync}. 
By contrast, a single ensemble of two-level constituents has only one local transition~($|0\rangle \leftrightarrow |1\rangle$), and hence only one corresponding coherence channel. Its internal mechanism for sustaining a oscillating phase is therefore much more restricted. This suggests a natural alternative rather than enlarging the local Hilbert space: couple two two-level ensembles so that the different ensembles provide multiple interacting coherence and population channels. Such a composite two-level system may reproduce, at the collective level, part of the dynamical structure that arises intrinsically within a single multilevel constituent~\cite{dosPrazeres2021DLevelBTC}.

Synchronization describes phase or frequency locking among preexisting oscillators~\cite{
	Kuramoto1975SelfEntrainment,Pikovsky2001Synchronization,Acebron2005KuramotoReview,
	Lee2013QuantumSync,Mari2013QuantumSync,Walter2014QuantumSync,
	Roulet2018QuantumSyncMeasures,Lee2014MeanField,Kato2025SpinVanDerPol,
	Lorenzo2022ChiralNetworks,Wadenpfuhl2023RydbergSync,Dai2026SpinSyncControl,Liu2026QuantumVdPSync}.
Related many-body settings include coupled oscillator groups and interacting time-crystalline components~\cite{
	Xu2014TwoEnsemblesAtoms,Nadolny2023MacroscopicSync,HongStrogatz2011ConformistContrarian,Sonnenschein2015TwoPopulations,
	Autti2021JosephsonTC,Autti2022NonlinearTwoLevelTC,
	Paulino2025Thermodynamics}, whose constituents typically oscillate before coupling.
Here, by contrast, we consider two locally pumped and dissipative spin-$1/2$ ensembles, neither of which forms an autonomous time crystal. 
Although collective dissipation can support time-crystalline order in a two-level ensemble~\cite{Iemini2018BoundaryTC,Tucker2018ShatteredTime}, local dissipation provides no such mechanism. 
Interensemble coupling instead generates a single collective oscillatory mode with a fixed relative phase—an interlocked temporal order rather than two independently formed time crystals that subsequently synchronize. 
This relative phase is analogous to the internal displacement within a multi-site spatial unit cell.
\begin{figure*}[ht] 
	\includegraphics[width=1\linewidth]{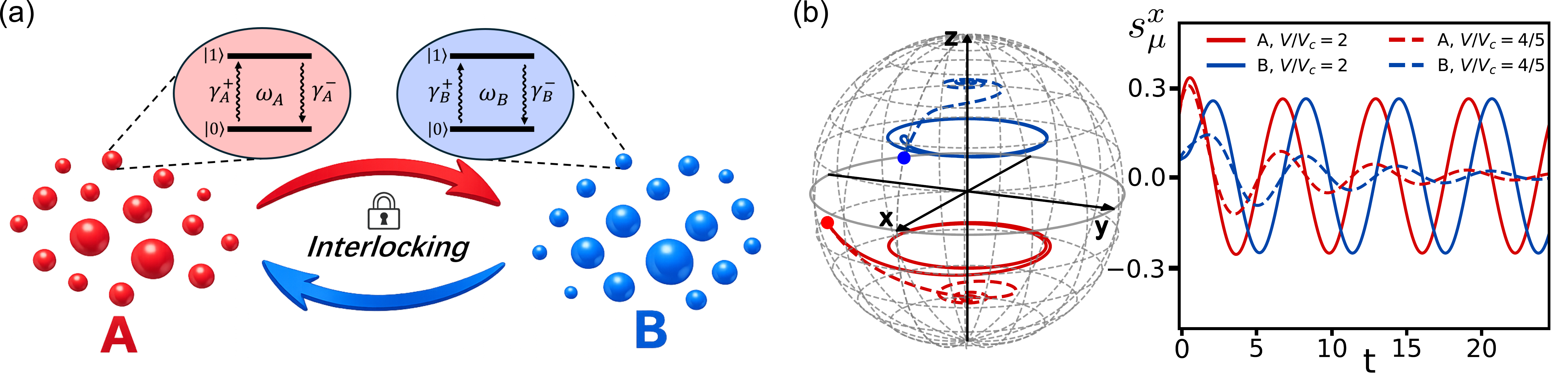}
	\caption{
		Interlocked dynamics of two dissipative ensembles.
		(a) Schematic illustration of two coupled dissipative spin ensembles A and B with incoherent pumping $\gamma_{\mu}^+$, decay $\gamma_{\mu}^-$, and bare transition frequencies $\omega_\mu$.
		(b) Mean-field dynamics. Left: trajectories of the normalized collective spins on the Bloch sphere. Dashed (solid) curves correspond to the stationary ($V<V_c$) [oscillatory ($V>V_c$)] phase. Right: time evolution of the transverse spin component, showing persistent oscillations above the critical interaction strength.
	}
	\label{fig1}
\end{figure*}

We establish this interlocked time-crystalline phase using mean-field theory, exact finite-size Liouvillian spectra, steady-state two-time correlations, and cumulant dynamics. 
Mean-field stability yields the collective frequency and critical coupling, while the exact spectra and correlations reveal a slow oscillatory mode whose lifetime grows with system size.
Cumulant dynamics shows that intra-ensemble order and inter-ensemble coherence share the same collective spectral branch. Together, these results identify a time-crystalline phase in which two individually non-oscillatory dissipative components form a single internally structured temporal order.

\textit{Model---}
We consider a pair of dissipative two-level ensembles A and B, each consisting of spin-1/2 particles [see Fig.~\ref{fig1}(a)], whose dynamics are governed by the collective spin operators $S^\alpha_\mu=\sum_{i=1}^{N}\sigma_{\mu,i}^\alpha$ with $\alpha=x,y,z$ and $\mu=A,B$. 
To simplify the analysis, intra-ensemble interactions are neglected, while all-to-all coupling between particles in different ensembles is retained and described by the Hamiltonian
\begin{equation}
	H=\omega_AS_A^z+\omega_BS_B^z+\frac{V}{N}\left(e^{i\theta}S_A^+S_B^-+H.c.\right)
\end{equation}
where $\omega_A$ and $\omega_B$ denote the bare transition frequencies of ensembles $A$ and $B$, respectively, $V$ characterizes the inter-ensemble coupling strength, and $\theta$ represents the relative phase associated with the interaction.
For simplicity, we focus on \(\theta=0\) in the main text; finite \(\theta\) shifts the locked relative phase without qualitatively changing the collective phase structure.
The more general case including intra-ensemble interactions is discussed the Supplemental Material(SM)~\cite{SM}, where we show that the resulting nonlinear frequency shift can be compensated by tuning the bare detuning.
Throughout this work, we adopt units in which $\hbar=1$.

In addition, each particle is subject to local dissipative processes, including incoherent pumping and decay. 
The system density matrix $\rho$ then evolves according to a Markovian master equation of Lindblad form~\cite{Gorini1976GKSL,Lindblad1976Generator}, $\partial_t\rho=\mathcal{L}(\rho)$, where the Liouvillian operator acts as 
\begin{equation}
	\mathcal{L}(\bullet)=-i[H,\bullet]+\sum_\mu\sum_i\left(\gamma_\mu^+\mathcal{D}[\sigma_{\mu,i}^+]+\gamma_\mu^-\mathcal{D}[\sigma_{\mu,i}^-]\right)\bullet,
\end{equation}
where $\gamma_\mu^\pm$ denote the incoherent pumping and decay rates of ensemble $\mu$, and $\mathcal{D}[O]\rho=O\rho O^\dagger-\frac{1}{2}\{O^\dagger O,\rho\}$ is the standard Lindblad dissipator.

\textit{Mean-Field Collective Instability---}
We first analyze the macroscopic dynamics at the mean-field level.
We introduce the normalized collective spin component $s_\mu^\alpha=\langle S_\mu^\alpha\rangle/N$ and the the transverse collective coherence $s_\mu^+=s_\mu^x+i s_\mu^y$.
For the $1/N$-scaled all-to-all inter-ensemble exchange considered here~\cite{Kac1963VanderWaals}, the mean-field factorization gives the exact macroscopic one-body dynamics in the thermodynamic limit for separable permutation-symmetric initial states~\cite{Mattes_PRL_2025,Carollo2024MeanFieldInfiniteRange,Spohn1980KineticEquations}.
Finite-size correlations and steady-state two-time functions are treated below.

\begin{figure}[t] 
\includegraphics[width=1\linewidth]{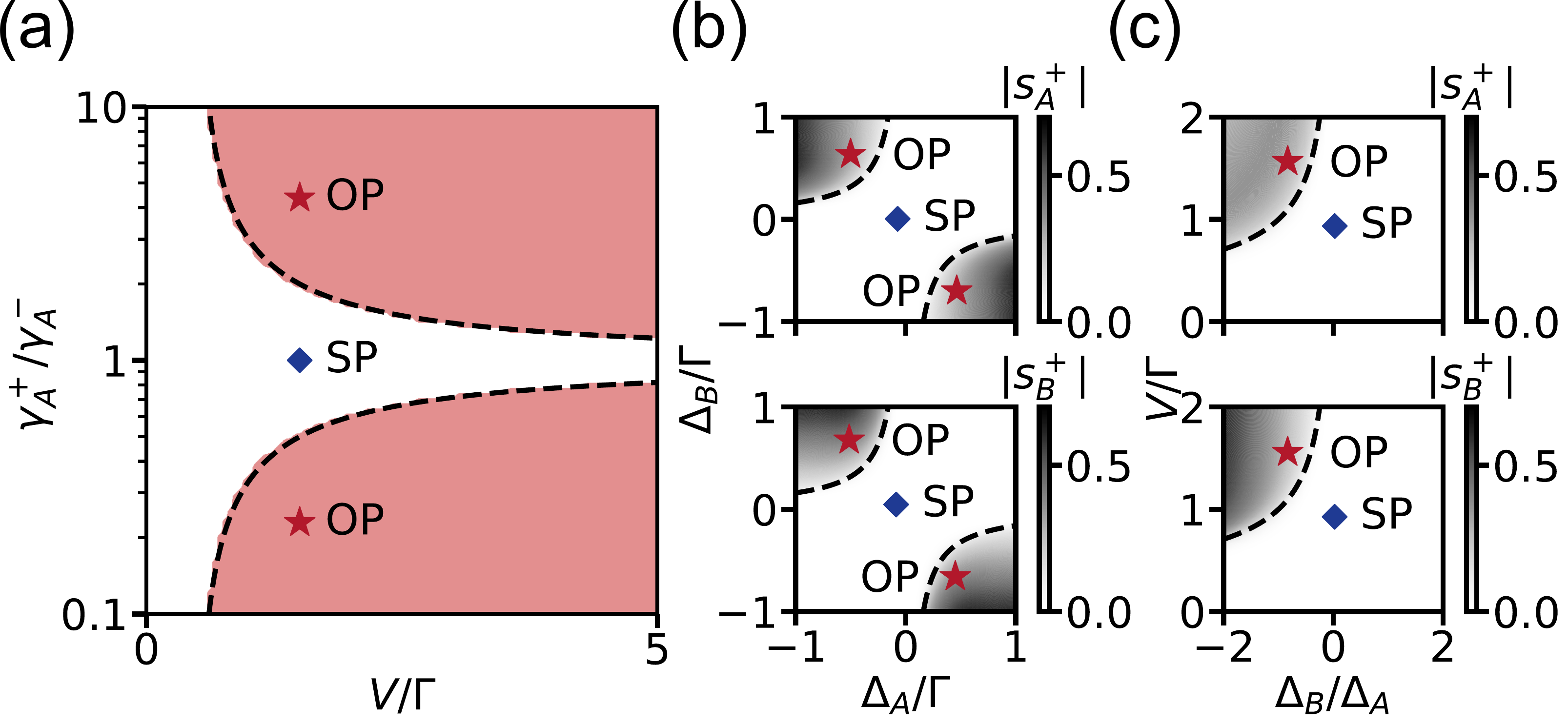}
\caption{
    Mean-field dynamical phase diagrams for the symmetric case.
    Phase diagrams are shown in the planes of
    (a) the interaction strength $V/\Gamma$ and pumping ratio $\gamma_A^+/\gamma_A^-$,
    (b) the nonequilibrium parameters $\Delta_A/\Gamma$ and $\Delta_B/\Gamma$,
    and (c) the interaction strength $V/\Gamma$ and imbalance ratio $\Delta_B/\Delta_A$.
    Dashed curves denote analytical phase boundaries obtained from the stability condition.
    White regions correspond to $|s_A^+|=|s_B^+|=0$, while shaded regions indicate finite transverse coherence in both ensembles.
    }
\label{fig2}
\end{figure}

We define the total local dissipation rates $\Gamma_\mu=\gamma_\mu^+ +\gamma_\mu^-$ and the pump-decay imbalances $\Delta_\mu=\gamma_\mu^+ -\gamma_\mu^-$, which quantify the imbalance between incoherent pumping and decay in ensemble $\mu$.
Numerical integration of the mean-field equations reveals two dynamical regimes [see Fig.~\ref{fig1}(b)].
For weak coupling, the system relaxes to a stationary phase (SP) with $s_{\mu,\mathrm{sp}}^z=\Delta_\mu/\Gamma_\mu$ and $s_{\mu,\mathrm{sp}}^+=0$.
Above a critical interaction strength, it enters an oscillatory phase (OP), characterized by persistent rotation of the transverse collective coherences.
We use $|s_\mu^+|_{t\to\infty}$ as the mean-field order parameter.
Figure~\ref{fig2} shows the mean-field dynamical phase diagrams for the symmetric case $\omega_A=\omega_B$, $\Gamma_A=\Gamma_B=\Gamma$, with $\gamma_A^+/\gamma_A^-=\gamma_B^-/\gamma_B^+$.
White regions correspond to the SP, where $|s_A^+|=|s_B^+|=0$, while shaded regions correspond to the OP, where both ensembles have finite transverse coherence.
No parameter regime is found in which only one ensemble develops an OP, indicating that the transition is collective.

Writing the transverse collective variables as $s_\mu^+=r_\mu e^{-i\phi_\mu}$, we further find that the relative phase $\Delta\phi=\phi_A-\phi_B$ approaches a constant value throughout the OP.
This locked phase difference gives the collective OP an internal phase structure, analogous to the fixed displacement of a two-site basis in a spatial crystal.
The analytical expression for $\Delta\phi$ and its dependence on the interaction phase $\theta$ are derived in the SM~\cite{SM}.

The onset of the OP is obtained by a mean-field stability analysis of the stationary phase, as commonly used to identify limit-cycle instabilities in dissipative many-body systems~\cite{Chan2015LimitCycle,Iemini2018BoundaryTC,Wang2025LocalDissipationBTC}.
To leading order in $s_A^+$ and $s_B^+$, one obtains
\begin{equation}
\frac{d}{dt}
\begin{pmatrix}
s_A^+ \\
s_B^+
\end{pmatrix}
=
\begin{pmatrix}
i\omega_A-\Gamma_A/2
&
-iV e^{-i\theta}s_{A,\mathrm{sp}}^z
\\
-iV e^{i\theta}s_{B,\mathrm{sp}}^z
&
i\omega_B-\Gamma_B/2
\end{pmatrix}
\begin{pmatrix}
s_A^+ \\
s_B^+
\end{pmatrix}.
\label{eq:linear_stability}
\end{equation}
The diagonal terms describe local transverse damping and bare precession, while the off-diagonal terms show that the coherent exchange is weighted by the stationary longitudinal polarizations prepared by local dissipation.
At the instability threshold, the critical eigenvalue of Eq.~\eqref{eq:linear_stability} becomes purely imaginary, $\lambda=i\omega_c$, yielding the collective oscillation frequency
\begin{equation}
\omega_c=\bar{\omega}
-\frac{\Delta\omega}{2}
\frac{\Gamma_A-\Gamma_B}{\Gamma_A+\Gamma_B},
\label{eq:collective_frequency}
\end{equation}
where the center frequency $\bar{\omega}=(\omega_A+\omega_B)/2$ and the frequency detuning $\Delta\omega=\omega_A-\omega_B$.
The onset of the OP is determined by the critical interaction strength
\begin{equation}
V_c^2=
-\frac{(\Gamma_A\Gamma_B)^2}{\Delta_A\Delta_B}
\left[
\frac{1}{4}
+\frac{\Delta\omega^2}{(\Gamma_A+\Gamma_B)^2}
\right].
\label{eq:critical_coupling}
\end{equation}
The condition $V_c^2>0$ immediately requires $\Delta_A\Delta_B<0$, showing that the two ensembles must possess opposite pump-decay imbalances in order to destabilize the SP and support the collective OP.
The resulting critical interaction strength determines the analytical phase boundaries shown as dashed curves in Fig.~\ref{fig2}.
The same critical eigenmode sets the locked relative phase, while the interaction phase $\theta$ shifts it without affecting the collective-instability condition.
Details of the phase dynamics, stability analysis, and detuning-dependent locking window are given in the SM~\cite{SM}.

Eq.~\eqref{eq:linear_stability} also clarifies the physical mechanism.
Local pumping and decay do not by themselves generate transverse order: they damp each transverse coherence at the rate $\Gamma_\mu/2$, while preparing the stationary longitudinal polarization $s_{\mu,\mathrm{sp}}^z=\Delta_\mu/\Gamma_\mu$.
The coherent exchange then closes a feedback loop between two otherwise damped transverse modes.
For opposite stationary polarizations, this loop converts exchange coupling into collective anti-damping and destabilizes the SP.
For equal-sign imbalances, it produces no positive growth rate, leaving the SP stable.
Thus the OP is generated by a collective instability of the coupled locally dissipative state, rather than by either ensemble alone or by phase locking two preexisting oscillators in the usual synchronization sense~\cite{Xu2014TwoEnsemblesAtoms,Nadolny2023MacroscopicSync,Arumugam2026InjectionLocking,Postavova2026CoupledDTCChaosSync}.

\textit{Evidence for Time-Crystalline Order---}
We now establish that the collective OP identified above corresponds to a genuine time-crystalline phase rather than a finite-size mean-field limit cycle. 
For finite \(N\) systems, the Liouvillian possesses a unique steady state, so spontaneous symmetry breaking is restored and temporal order must instead be diagnosed through long-lived Liouvillian modes, closing Liouvillian gaps, and steady-state two-time correlations~\cite{Minganti2018SpectralTheory,Iemini2018BoundaryTC,Lledo2020DissipativeTC,Booker2020Nonstationarity,Souza2023GaplessLindbladians}.
\begin{figure}[b]
	\includegraphics[width=1\linewidth]{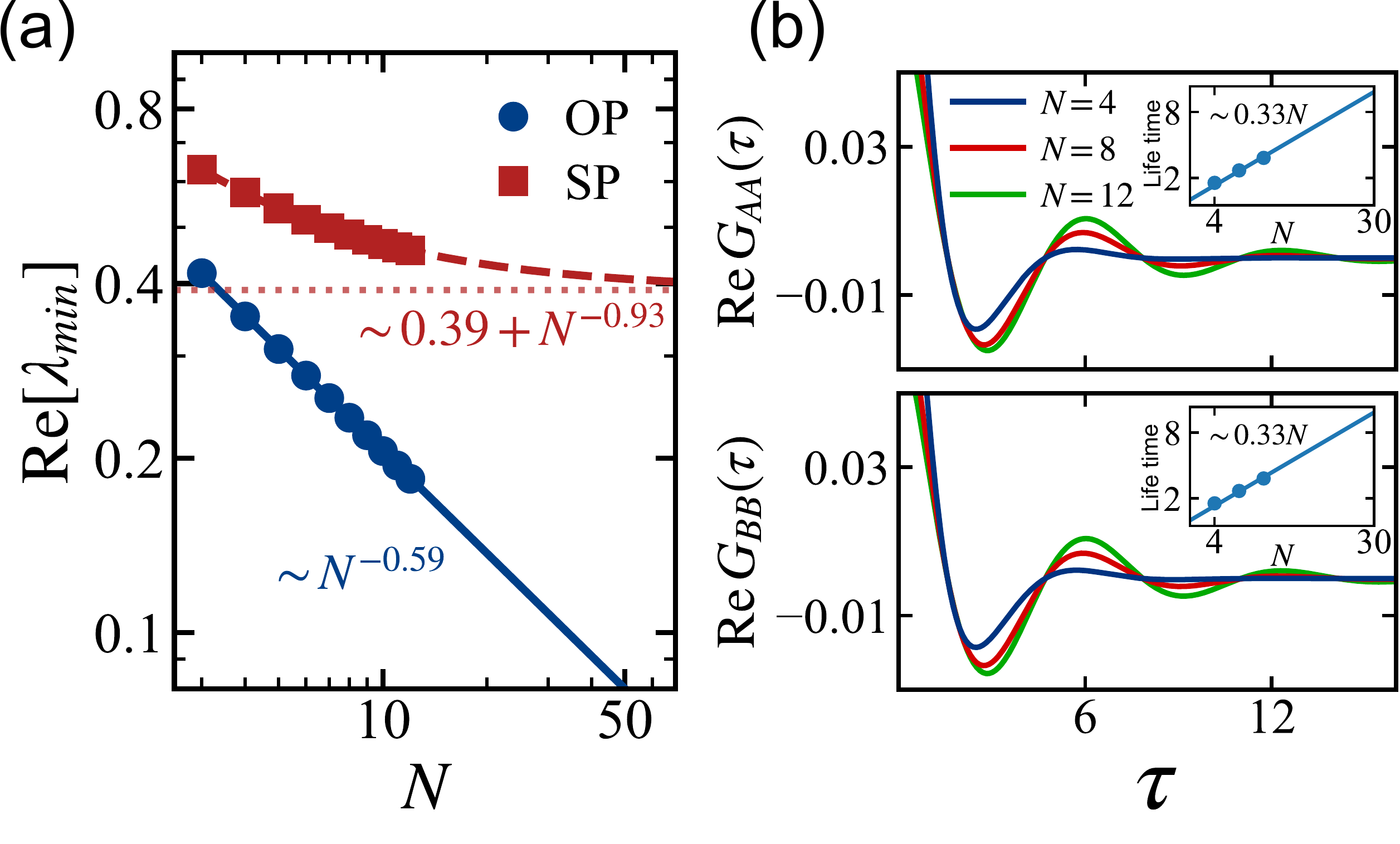}
	\caption{
		Liouvillian gap and two-time correlation scaling.
		(a) Scaling of the decay rate \(-\mathrm{Re}[\lambda_{\rm slow}]\) in the OP and SP.
		The OP exhibits algebraic gap closing, while the SP approaches a finite decay rate.
		(b) Real parts of \(G_{AA}(\tau)\) and \(G_{BB}(\tau)\) in the OP.
		Insets show lifetimes extracted from exponential fits.
		Parameters are \(\gamma_A^+=\gamma_B^-=0.6\), \(\gamma_A^-=\gamma_B^+=0.2\), and \(V=1\).
	}
	\label{fig3}
\end{figure}
Using the permutation symmetry within each ensemble, we diagonalize the finite-size Liouvillian in the symmetric Dicke representation~\cite{Dicke1954Coherence,Shammah2018PIQS} and identify the slowest decaying nonsteady mode \(\lambda_{\rm slow}\),
\begin{equation}
	\mathrm{Re}[\lambda_{\rm slow}]
	=
	\max_{k\neq0}\mathrm{Re}[\lambda_k],
	\qquad
	\mathcal L(\rho_k)=\lambda_k\rho_k .
\end{equation}
We also compute the normalized steady-state two-time correlations
\begin{equation}
	G_{\mu\nu}(\tau)
	=
	\langle S_\mu^+(\tau)S_\nu^-(0)\rangle_{\rm ss}/N^2,
\end{equation}
using the quantum regression theorem~\cite{Lax1963QuantumRegression,GardinerZoller2004QuantumNoise}.

Figure~\ref{fig3}(a) shows that the OP decay rate closes algebraically with system size,
\(-\mathrm{Re}[\lambda_{\rm slow}]\propto N^{-c}\), with \(c\simeq0.59\) for the parameters considered here. 
In contrast, in the SP the decay rate approaches a finite value
\(-\mathrm{Re}[\lambda_{\rm slow}]\to \kappa_\infty\simeq0.4\), consistent with
the intrinsic transverse relaxation rate of the SP, \(\Gamma/2\), under the present parameters.
The same distinction is visible in the two-time correlations: as shown in Fig.~\ref{fig3}(b), both \(G_{AA}(\tau)\) and \(G_{BB}(\tau)\) develop increasingly persistent oscillations, and their fitted lifetimes grow with \(N\). 
Thus the finite-size damping of the oscillatory mode vanishes in the thermodynamic limit, establishing persistent temporal order in both ensembles.
These results establish time-crystalline order, but they do not yet determine whether the two ensembles represent two locked temporal orders or a single interlocked one.

To distinguish this phase from two independently formed time crystals that are later phase locked, we next examine the internal correlation structuree.
We introduce the connected transverse correlations~\cite{Lourenco2022MultipartiteBTC}
\begin{equation}
    C_{\mu\nu}^{+-}
    =
    \langle\sigma_{\mu,i}^{+}\sigma_{\nu,j}^{-}\rangle
    -
    \langle\sigma_{\mu,i}^{+}\rangle
    \langle\sigma_{\nu,j}^{-}\rangle,
\end{equation}
with \(i\neq j\) for \(\mu=\nu\). 
Within the second-order cumulant expansion~\cite{Kubo1962CumulantExpansion,Plankensteiner2022QuantumCumulants}, the vector
\(\boldsymbol{\mathcal C}=\left(C_{AA}^{+-},C_{BB}^{+-},C_{AB}^{+-},C_{AB}^{-+}\right)^T\)
obeys \(\dot{\boldsymbol{\mathcal C}}=M_c\boldsymbol{\mathcal C}+\mathbf b_c\), where
\begin{equation}
	M_c=
	\begin{pmatrix}
		M_{\rm intra} & M_{12}\\
		M_{21} & M_{\rm inter}
	\end{pmatrix},
	\\
	M_{\rm intra}=
	\begin{pmatrix}
		-\Gamma_A & 0\\
		0 & -\Gamma_B
	\end{pmatrix}.
	\label{eq:Mc}
\end{equation}
The remaining blocks are given in the SM~\cite{SM}. 

Eq.~\eqref{eq:Mc} shows that, without the off-diagonal blocks \(M_{12}\) and \(M_{21}\), the intra-ensemble correlations are purely damped. 
Finite long-time correlations within either ensemble therefore require dynamical coupling to the inter-ensemble correlation sector, providing a correlation-level signature of interlocking.
We now test this structural implication directly by evaluating the steady-state correlations and their spectra.
\begin{figure}[t]
	\includegraphics[width=1\linewidth]{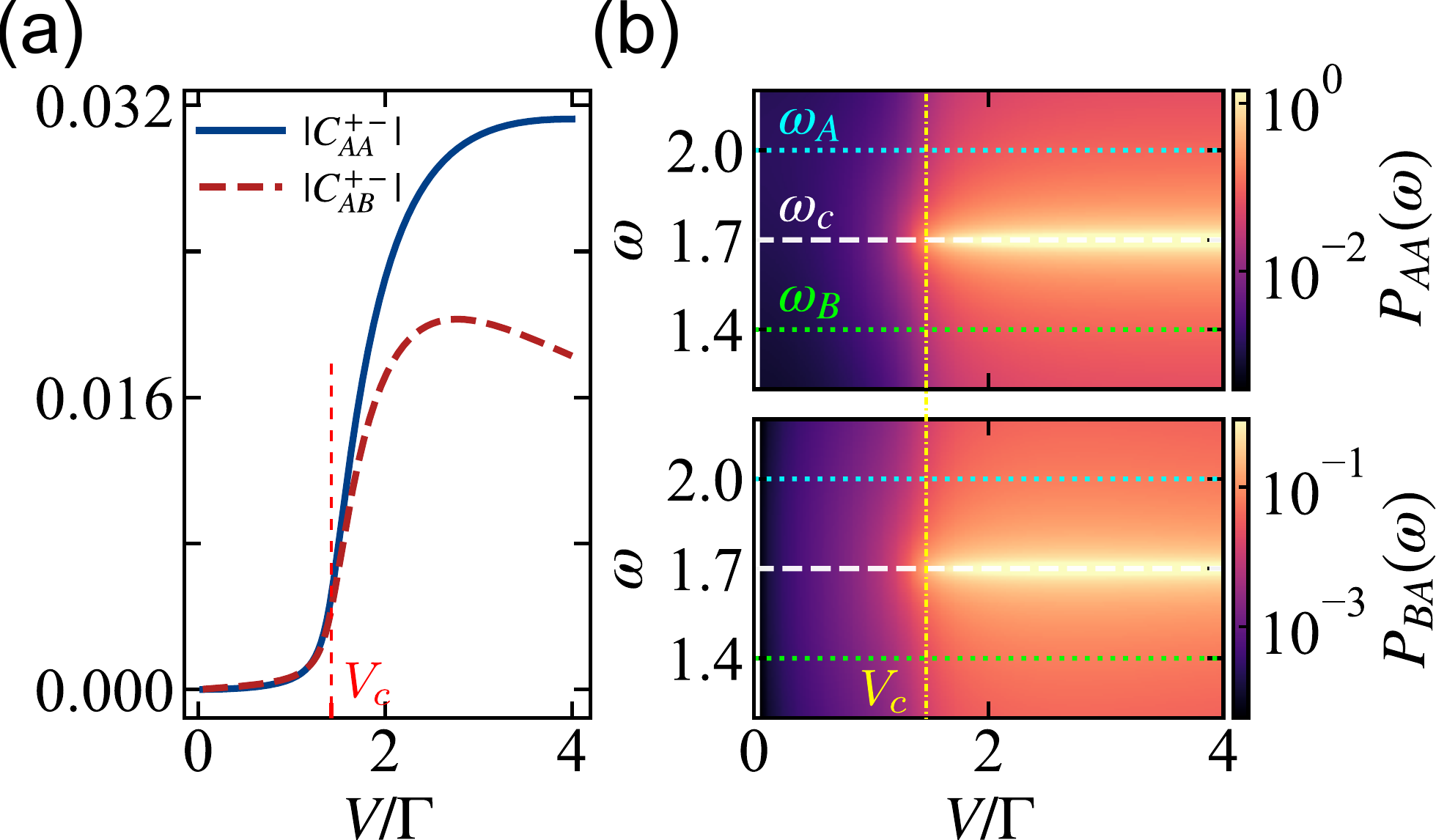}
	\caption{
		Correlation signatures of the interlocked time-crystal phase from the cumulant expansion.
		(a) Equal-time connected correlations \(|C_{AA}^{+-}|\) and \(|C_{AB}^{+-}|\) versus \(V/\Gamma\).
		(b) Frequency-resolved spectra \(P_{AA}(\omega)\) and \(P_{BA}(\omega)\) for \(\omega_A\neq\omega_B\).
		Vertical lines mark \(V_c\); horizontal guide lines indicate \(\omega_A\), \(\omega_B\), and \(\omega_c\).
		Calculations are performed for \(N=1000\).
	}
	\label{fig4}
\end{figure}

This structure is confirmed in Fig.~\ref{fig4}(a), where the intra-ensemble correlation \(|C_{AA}^{+-}|\) and inter-ensemble correlation \(|C_{AB}^{+-}|\) emerge together at the transition. 
The small finite values below \(V_c\) reflect finite-size rounding, while the correlated growth above \(V_c\) shows that intra-ensemble coherence is built together with inter-ensemble coherence. 
To resolve the temporal content of these correlations, we compute the spectra
\begin{equation}
    P_{\mu\nu}(\omega)=\left|\int_0^\infty d\tau\,e^{i\omega\tau}G_{\mu\nu}(\tau)\right|.
\end{equation}
For the detuned case \(\omega_A\neq\omega_B\), Fig.~\ref{fig4}(b) shows that the diagonal spectrum \(P_{AA}(\omega)\) and the cross spectrum \(P_{BA}(\omega)\) select the same collective branch at \(\omega_c\), rather than separate branches near \(\omega_A\) and \(\omega_B\). 
Hence the temporal order within an ensemble and the temporal coherence between ensembles are governed by a single collective oscillatory mode.

Taken together, the Liouvillian gap closing and two-time correlations establish time-crystalline order, while the cumulant hierarchy and spectra show that this order is internally interlocked. 
The OP therefore has no single-ensemble counterpart and is not the phase locking of two preexisting time crystals, but a single time-crystalline phase with a two-component internal structure.

\textit{Robustness---}
We finally examine the robustness of the interlocked phase against experimentally relevant perturbations.
The analytical instability condition already indicates that the interlocked OP is not restricted to exact resonance.
A finite bare detuning \(\Delta\omega\) enters the critical coupling only through the positive correction
\(\Delta\omega^2/(\Gamma_A+\Gamma_B)^2\), increasing the required coupling without changing the collective-instability mechanism as long as \(\Delta_A\Delta_B<0\).
Consistently, the frequency-resolved spectra in the SM~\cite{SM} show a finite detuning window in which both ensembles select the same collective branch.
Outside this window, the collective peak is suppressed rather than splitting into two independent single-ensemble branches.
Thus this window should be understood as the detuning tolerance of a single interlocked OP, not as a synchronization region between two autonomous oscillators.

We further examine the effects of random white noise added to the inter-ensemble coupling and the local dissipation rates in the SM~\cite{SM}, showing that the interlocked phase remains robust against moderate stochastic fluctuations, as expected for a stable dissipative time-crystalline phase~\cite{Kongkhambut2022ContinuousTC,Wu2024RydbergDTC}.
This is important experimentally because the coupling, gain, and loss rates will generally fluctuate in realistic platforms.

Another unavoidable imperfection is residual intra-ensemble interaction, especially in cavity- or resonator-mediated settings where the same mode can mediate both inter- and intra-ensemble couplings~\cite{Sorensen2002BadCavities,Norcia2018CavitySpinExchange,Li2022CollectiveSpinLight}.
Such terms mainly renormalize the effective detuning, \(\Delta\omega\to\Delta\omega_{\rm eff}\), rather than generating an independent OP.
Related interaction-induced frequency shifts have also been identified in driven-dissipative collective spin systems~\cite{Nadolny2025SuperradiantFrequencyShift}.
In the SM~\cite{SM}, we further analyze the effect of residual intra-ensemble interactions and show that they primarily renormalize the effective detuning. They can therefore be treated as part of the detuning budget and compensated by tuning the bare transition frequencies, for example through differential Zeeman shifts, Stark shifts, or cavity detuning. 
These results show that the interlocked phase is robust against realistic imperfections and that residual intra-ensemble interactions can be compensated through effective detuning control.
The relevant experimental requirement is thus to keep the compensated effective detuning inside the collective interlocking window, rather than to eliminate all residual intra-ensemble interactions.

\textit{Conclusion and outlook---}
We have demonstrated an interlocked time crystal in which two dissipative ensembles form a single collective temporal phase with a fixed internal phase relation.
The key point is not merely that the two ensembles oscillate at the same frequency, but that neither ensemble supports an autonomous OP: the temporal order, inter-ensemble coherence, and collective spectral branch arise from the coupled dynamics as one structured phase.
In this sense, the locked relative phase plays the role of an internal temporal displacement, providing a time-domain analogue of an ordered structure with a nontrivial internal basis.

Beyond the specific spin-$1/2$ realization studied here, our results suggest that the absence of an autonomous order parameter in each dissipative subsystem does not preclude the emergence of time-crystalline order. 
Instead, the order parameter itself can be a genuinely collective object generated by the interaction between local dissipative channels. 
This perspective provides a unified interpretation of locally dissipative time-crystalline phases, in which additional internal levels or coupled subsystems supply the interacting dissipative channels required to generate the collective instability~\cite{Russo2025QDCTC,Wang2025LocalDissipationBTC}. 
When these dissipative channels are implemented as coupled subsystems rather than internal levels of a single unit, their mutual interactions become independently engineerable, connecting the present mechanism to broader studies of interacting time crystals and coupled active-spin ensembles~\cite{Autti2021JosephsonTC,Nadolny2025NonreciprocalSync,Paulino2025Thermodynamics}. 
The interaction phase therefore provides a control knob for the internal temporal structure of the order parameter. 
Extending this modular architecture to multiple locally dissipative components could realize multi-site temporal unit cells with programmable internal phase relations, opening a route toward time crystals with richer internal structure, in line with recent studies of coupled, clustered, and multiple time-crystalline phases~\cite{Solanki2024ExoticSync,Liu2025BifurcationTC,Jiao2025MultipleTC}. 
Such programmable interlocked temporal structures may also be useful for engineered nonequilibrium quantum devices, including quantum energy-storage protocols~\cite{Campaioli2024QuantumBatteries}, where coupled time crystals have recently been proposed as quantum batteries~\cite{Paulino2025Thermodynamics}.

\textit{Acknowledgments---}
	This work was supported by the National Natural Science Foundation of China(Grant No.12375013, No.12547109, No.92365109, and No.12474501), and Guangdong Provincial Quantum Science Strategic Initiative (Grant No. GDZX2503008).
    
\normalem
\bibliography{ref}


\clearpage
\onecolumngrid

\setcounter{page}{1}
\renewcommand{\thepage}{S\arabic{page}}

\setcounter{section}{0}
\renewcommand{\thesection}{S\arabic{section}}

\setcounter{equation}{0}
\renewcommand{\theequation}{S\arabic{equation}}

\setcounter{figure}{0}
\renewcommand{\thefigure}{S\arabic{figure}}

\setcounter{table}{0}
\renewcommand{\thetable}{S\arabic{table}}

\makeatletter
\setcounter{secnumdepth}{3}
\renewcommand{\thesection}{\arabic{section}}
\renewcommand{\thesubsection}{\Alph{subsection}}
\renewcommand{\citenumfont}[1]{S#1}
\renewcommand{\bibnumfmt}[1]{[S#1]}
\makeatother

\begin{center}
{\large\bf Supplemental Material for\\
``Interlocked Time Crystal in Coupled Spin-1/2 Ensembles under Local Dissipation''}\\[1ex]

Zhen-Huan Yang, Zhen-Tao Liang and Dan-Bo Zhang
\end{center}

	\section{Stability analysis and physical mechanism}
	\label{sec:SM_stability_mechanism}
	
	In this section we analyze the mean-field origin of the interlocked oscillatory phase.
	The purpose is twofold. First, we show explicitly that the stationary phase loses stability
	through a transverse Hopf instability, a standard route to limit-cycle dynamics in
	driven-dissipative systems~\cite{Chan2015LimitCycle,Buca2019Dissipation,Wang2025LocalDissipationBTC,Russo2025QDCTC}.
	Second, we clarify the physical feedback mechanism that makes this instability possible
	in two coupled two-level ensembles, while it is absent for a single self-coupled
	two-level ensemble with the same type of local gain and loss.
    This absence of an autonomous transverse instability in a single locally pumped and decaying two-level ensemble contrasts with multilevel local-dissipation mechanisms, where distinct internal coherence channels can provide the feedback needed for nonstationary order~\cite{Nadolny2023MacroscopicSync,Wang2025LocalDissipationBTC,Russo2025QDCTC}.
	
	\subsection{Mean-field equations and stationary phase}
	
	Starting from the master equation in the main text, we introduce the normalized collective
	variables
	\begin{equation}
		s^\alpha_\mu=\frac{\langle S^\alpha_\mu\rangle}{N},
		\qquad
		s^\pm_\mu=s^x_\mu\pm i s^y_\mu ,
		\qquad \mu=A,B .
	\end{equation}
	For the infinite-range interaction considered here, the mean-field factorization
	\begin{equation}
		\langle O_i O_j\rangle\simeq \langle O_i\rangle\langle O_j\rangle
	\end{equation}
	becomes exact in the thermodynamic limit for Kac-scaled infinite-range interactions~\cite{Kac1963VanderWaals,Nadolny2023MacroscopicSync,Wang2025LocalDissipationBTC}. The resulting mean-field equations are
	\begin{align}
		\dot s_A^+
		&=
		\left(i\omega_A-\frac{\Gamma_A}{2}\right)s_A^+
		-iV e^{-i\theta}s_A^z s_B^+ ,
		\label{eq:SM_splusA}
		\\
		\dot s_B^+
		&=
		\left(i\omega_B-\frac{\Gamma_B}{2}\right)s_B^+
		-iV e^{i\theta}s_B^z s_A^+ ,
		\label{eq:SM_splusB}
		\\
		\dot s_A^z
		&=
		-\Gamma_A s_A^z+\Delta_A
		+2iV
		\left(
		e^{i\theta}s_A^+s_B^-
		-
		e^{-i\theta}s_A^-s_B^+
		\right),
		\label{eq:SM_szA}
		\\
		\dot s_B^z
		&=
		-\Gamma_B s_B^z+\Delta_B
		-2iV
		\left(
		e^{i\theta}s_A^+s_B^-
		-
		e^{-i\theta}s_A^-s_B^+
		\right),
		\label{eq:SM_szB}
	\end{align}
	where
	\begin{equation}
		\Gamma_\mu=\gamma_{\mu,+}+\gamma_{\mu,-},
		\qquad
		\Delta_\mu=\gamma_{\mu,+}-\gamma_{\mu,-}.
	\end{equation}
	The stationary phase is characterized by vanishing transverse coherence,
	\begin{equation}
		s_A^+=s_B^+=0 ,
	\end{equation}
	and longitudinal components fixed by the local pump-decay imbalance,
	\begin{equation}
		s^z_{A,\mathrm{sp}}=\frac{\Delta_A}{\Gamma_A},
		\qquad
		s^z_{B,\mathrm{sp}}=\frac{\Delta_B}{\Gamma_B}.
		\label{eq:SM_sp_z}
	\end{equation}
	Thus each ensemble alone relaxes to an incoherent stationary state. The oscillatory phase
	can only arise if this stationary point becomes unstable in the transverse sector.
	
	\subsection{Stability analysis and locked eigenmode}
	
	We now linearize Eqs.~\eqref{eq:SM_splusA}--\eqref{eq:SM_szB} around the stationary phase.
	Because the transverse variables vanish at the stationary point, the longitudinal
	perturbations decouple to linear order and obey
	\begin{equation}
		\frac{d}{dt}
		\begin{pmatrix}
			\delta s_A^z\\
			\delta s_B^z
		\end{pmatrix}
		=
		\begin{pmatrix}
			-\Gamma_A & 0\\
			0 & -\Gamma_B
		\end{pmatrix}
		\begin{pmatrix}
			\delta s_A^z\\
			\delta s_B^z
		\end{pmatrix}.
	\end{equation}
	The longitudinal eigenvalues are therefore always stable. The instability of the
	stationary phase is determined by the transverse Jacobian. In the complex transverse
	representation
	\begin{equation}
		\delta\mathbf{s}_\perp=
		\begin{pmatrix}
			\delta s_A^+\\
			\delta s_B^+
		\end{pmatrix},
	\end{equation}
	the linearized dynamics is
	\begin{equation}
		\frac{d}{dt}\delta\mathbf{s}_\perp
		=
		M_\perp \delta\mathbf{s}_\perp ,
		\label{eq:SM_transverse_linear}
	\end{equation}
	with
	\begin{equation}
		M_\perp
		=
		\begin{pmatrix}
			i\omega_A-\Gamma_A/2
			&
			-iV e^{-i\theta}s^z_{A,\mathrm{sp}}
			\\
			-iV e^{i\theta}s^z_{B,\mathrm{sp}}
			&
			i\omega_B-\Gamma_B/2
		\end{pmatrix}.
		\label{eq:SM_transverse_matrix}
	\end{equation}
	This matrix is the transverse block of the Jacobian written in complex variables. The
	corresponding equations for $\delta s_A^-$ and $\delta s_B^-$ are obtained by complex
	conjugation and give the conjugate pair of eigenvalues. Hence a Hopf bifurcation occurs
    when an eigenvalue of $M_\perp$ crosses the imaginary axis, as in standard
    driven-dissipative limit-cycle transitions~\cite{Chan2015LimitCycle,Wang2025LocalDissipationBTC}.
	
	The characteristic equation is
	\begin{equation}
		\left(
		\lambda+\frac{\Gamma_A}{2}-i\omega_A
		\right)
		\left(
		\lambda+\frac{\Gamma_B}{2}-i\omega_B
		\right)
		+
		V^2s^z_{A,\mathrm{sp}}s^z_{B,\mathrm{sp}}
		=0.
		\label{eq:SM_char_eq}
	\end{equation}
	At the instability threshold we set
	\begin{equation}
		\lambda=i\omega_c ,
	\end{equation}
	where $\omega_c$ is the collective oscillation frequency at the onset of the oscillatory
	phase. Separating Eq.~\eqref{eq:SM_char_eq} into real and imaginary parts gives
	\begin{equation}
		\frac{\Gamma_A}{2}(\omega_c-\omega_B)
		+
		\frac{\Gamma_B}{2}(\omega_c-\omega_A)
		=0 .
	\end{equation}
	Thus
	\begin{equation}
		\omega_c
		=
		\bar\omega
		-
		\frac{\Delta\omega}{2}
		\frac{\Gamma_A-\Gamma_B}{\Gamma_A+\Gamma_B},
		\label{eq:SM_omega_c}
	\end{equation}
	where
	\begin{equation}
		\bar\omega=\frac{\omega_A+\omega_B}{2},
		\qquad
		\Delta\omega=\omega_A-\omega_B .
	\end{equation}
	The real part of Eq.~\eqref{eq:SM_char_eq} then yields the critical coupling
	\begin{equation}
		V_c^2
		=
		-
		\frac{(\Gamma_A\Gamma_B)^2}{\Delta_A\Delta_B}
		\left[
		\frac{1}{4}
		+
		\frac{\Delta\omega^2}{(\Gamma_A+\Gamma_B)^2}
		\right].
		\label{eq:SM_Vc}
	\end{equation}
	Therefore the condition $V_c^2>0$ requires
	\begin{equation}
		\Delta_A\Delta_B<0.
		\label{eq:SM_opposite_imbalance}
	\end{equation}
	The two ensembles must have opposite pump-decay imbalance in order for the stationary
	phase to lose transverse stability.
	
	The same critical eigenmode also determines the locked internal phase. Writing the
	unstable eigenmode as
	\begin{equation}
		\delta s_A^+(t)=a_A e^{i\omega_c t},
		\qquad
		\delta s_B^+(t)=a_B e^{i\omega_c t},
	\end{equation}
	and using the convention
	\begin{equation}
		s_\mu^+=r_\mu e^{-i\phi_\mu},
	\end{equation}
	the relative phase is
	\begin{equation}
		\Delta\phi=\phi_A-\phi_B
		=
		\arg\left(\frac{a_B}{a_A}\right).
		\label{eq:SM_phase_from_eigenmode}
	\end{equation}
	From the first row of the eigenvalue equation
	\begin{equation}
		(M_\perp-i\omega_c I)
		\begin{pmatrix}
			a_A\\
			a_B
		\end{pmatrix}
		=0,
	\end{equation}
	we obtain
	\begin{equation}
		\frac{a_B}{a_A}
		=
		\frac{e^{i\theta}}{V s^z_{A,\mathrm{sp}}}
		\left(
		\omega_A-\omega_c+\frac{i\Gamma_A}{2}
		\right).
	\end{equation}
	Therefore
	\begin{equation}
		\Delta\phi
		=
		\theta
		+
		\arg\left(
		\omega_A-\omega_c+\frac{i\Gamma_A}{2}
		\right)
		-
		\arg(s^z_{A,\mathrm{sp}}).
		\label{eq:SM_locked_phase_A}
	\end{equation}
	Equivalently, using
	\begin{equation}
		\omega_A-\omega_c
		=
		\frac{\Gamma_A}{\Gamma_A+\Gamma_B}\Delta\omega ,
	\end{equation}
	one may write
	\begin{equation}
		\Delta\phi
		=
		\theta
		+
		\arg\left[
		2\Delta\omega+i(\Gamma_A+\Gamma_B)
		\right]
		-
		\arg(s^z_{A,\mathrm{sp}}).
		\label{eq:SM_locked_phase_A2}
	\end{equation}
	The second row gives the equivalent expression
	\begin{equation}
		\Delta\phi
		=
		\theta
		+
		\arg(s^z_{B,\mathrm{sp}})
		-
		\arg\left[
		-2\Delta\omega+i(\Gamma_A+\Gamma_B)
		\right],
		\label{eq:SM_locked_phase_B}
	\end{equation}
	with the two forms being equivalent modulo $2\pi$ when
	$s^z_{A,\mathrm{sp}}s^z_{B,\mathrm{sp}}<0$.
	
	Thus the oscillation frequency and the internal phase relation are fixed by the same
	critical eigenmode. The transition is therefore not the synchronization of two preexisting
	oscillators in the usual sense of coupled autonomous oscillators~\cite{Kuramoto1975SelfEntrainment,Pikovsky2001Synchronization,Acebron2005KuramotoReview,Nadolny2023MacroscopicSync}.
	Instead, it is a Hopf instability of a single collective transverse mode of the
	coupled stationary state.
	
	\subsection{Physical feedback mechanism}
	
	To clarify why the interlocked oscillatory phase requires cross-feedback between two
	ensembles, we first compare it with a single two-level ensemble subject to the same type
	of local incoherent pumping, decay, and intra-ensemble exchange interaction. The
	single-ensemble Hamiltonian is
	\begin{equation}
		H_{\rm single}
		=
		\omega S^z
		+
		\frac{V_{\rm intra}}{N}
		\sum_{i<j}
		\left(
		\sigma_i^+\sigma_j^-
		+
		\sigma_i^-\sigma_j^+
		\right).
		\label{eq:SM_H_single}
	\end{equation}
	The density matrix evolves according to
	\begin{equation}
		\dot\rho
		=
		-i[H_{\rm single},\rho]
		+
		\sum_{i=1}^{N}
		\left(
		\gamma_+\mathcal D[\sigma_i^+]\rho
		+
		\gamma_-\mathcal D[\sigma_i^-]\rho
		\right),
		\label{eq:SM_single_master}
	\end{equation}
	where $\mathcal D[O]\rho=O\rho O^\dagger-\{O^\dagger O,\rho\}/2$.
	We also define
	\begin{equation}
		\Gamma=\gamma_+ + \gamma_-,
		\qquad
		\Delta=\gamma_+ - \gamma_- .
	\end{equation}
	At the mean-field level, Eq.~\eqref{eq:SM_H_single} gives
	\begin{equation}
		\dot s^+
		=
		\left[
		i\omega
		-
		\frac{\Gamma}{2}
		-
		i\frac{N-1}{N}V_{\rm intra}s^z
		\right]s^+ ,
		\label{eq:SM_single_splus}
	\end{equation}
	and
	\begin{equation}
		\dot s^z
		=
		-\Gamma s^z+\Delta .
		\label{eq:SM_single_sz}
	\end{equation}
	The factor $(N-1)/N$ appears because each spin interacts with the other $N-1$ spins in
	the same ensemble. In the thermodynamic limit,
	\begin{equation}
		\frac{N-1}{N}\rightarrow 1,
	\end{equation}
	so this factor does not change the qualitative conclusion.
	
	The stationary state of the single ensemble is
	\begin{equation}
		s^+_{sp,*}=0,
		\qquad
		s^z_{sp,*}=\frac{\Delta}{\Gamma}.
	\end{equation}
	Linearizing Eq.~\eqref{eq:SM_single_splus} around this stationary state yields
	\begin{equation}
		\lambda_{\rm single}
		=
		-\frac{\Gamma}{2}
		+
		i\left(
		\omega
		-
		\frac{N-1}{N}
		V_{\rm intra}
		\frac{\Delta}{\Gamma}
		\right).
		\label{eq:SM_single_lambda}
	\end{equation}
	Therefore
	\begin{equation}
		Re[\lambda_{\rm single}]
		=
		-\frac{\Gamma}{2}<0 .
	\end{equation}
	The intra-ensemble exchange interaction only renormalizes the precession frequency,
	\begin{equation}
		\omega
		\rightarrow
		\omega_{\rm eff}
		=
		\omega
		-
		\frac{N-1}{N}
		V_{\rm intra}
		\frac{\Delta}{\Gamma},
		\label{eq:SM_single_omega_eff}
	\end{equation}
	which becomes
	\begin{equation}
		\omega_{\rm eff}
		\rightarrow
		\omega
		-
		V_{\rm intra}
		\frac{\Delta}{\Gamma}
	\end{equation}
	for $N\rightarrow\infty$. Thus self-feedback in a single two-level ensemble is reactive:
	it shifts the oscillation frequency but does not modify the transverse damping rate. It
	therefore cannot destabilize the stationary phase or generate an autonomous oscillatory
	phase within this local gain-loss and exchange-feedback mechanism.
	
	The coupled two-ensemble system is qualitatively different. The inter-ensemble exchange
	interaction does not act as a self-frequency shift of one transverse coherence. Instead, it
	generates a cross-feedback loop between the two coherences,
	\begin{equation}
		s_A^+
		\longrightarrow
		s_B^+
		\longrightarrow
		s_A^+ .
	\end{equation}
	This feedback loop is already visible in the transverse stability matrix,
	\begin{equation}
		M_\perp
		=
		\begin{pmatrix}
			i\omega_A-\Gamma_A/2
			&
			-iV e^{-i\theta}s^z_{A,\mathrm{sp}}
			\\
			-iV e^{i\theta}s^z_{B,\mathrm{sp}}
			&
			i\omega_B-\Gamma_B/2
		\end{pmatrix}.
		\label{eq:SM_feedback_matrix}
	\end{equation}
	One round of the feedback loop carries the factor
	\begin{equation}
		(-iVs_A^z)(-iVs_B^z)
		=
		-V^2s_A^z s_B^z .
		\label{eq:SM_feedback_loop}
	\end{equation}
	If
	\begin{equation}
		s_A^z s_B^z>0,
	\end{equation}
	then this feedback is effectively reactive and mainly produces frequency hybridization.
	For example, in the resonant symmetric case
	\begin{equation}
		\omega_A=\omega_B=\omega,
		\qquad
		\Gamma_A=\Gamma_B=\Gamma,
	\end{equation}
	the transverse eigenvalues are
	\begin{equation}
		\lambda_\pm
		=
		i\omega-\frac{\Gamma}{2}
		\pm
		V\sqrt{-s^z_{A,\mathrm{sp}}s^z_{B,\mathrm{sp}}}.
		\label{eq:SM_symmetric_eigenvalues}
	\end{equation}
	When $s^z_{A,\mathrm{sp}}s^z_{B,\mathrm{sp}}>0$, the square root is imaginary, so the
	interaction only splits the oscillation frequencies while the real part remains
	$-\Gamma/2$.
	
	By contrast, if
	\begin{equation}
		s_A^z s_B^z<0,
	\end{equation}
	then
	\begin{equation}
		-V^2s_A^z s_B^z>0 .
	\end{equation}
	The cross-feedback loop produces a real eigenvalue splitting of the transverse Jacobian.
	One collective transverse mode is shifted toward positive real part and can overcome the
	local damping. This is the physical origin of the condition
	\begin{equation}
		\Delta_A\Delta_B<0,
	\end{equation}
	because
	\begin{equation}
		s^z_{A,\mathrm{sp}}s^z_{B,\mathrm{sp}}
		=
		\frac{\Delta_A\Delta_B}{\Gamma_A\Gamma_B}.
	\end{equation}
	Thus opposite pump-decay imbalance prepares opposite longitudinal polarizations, and
	these opposite polarizations convert coherent inter-ensemble exchange into collective
	anti-damping.
	
	The same feedback picture also explains why the oscillatory phase selects a fixed
	internal phase beyond the linear instability. To see this, we write
	\begin{equation}
		s_\mu^+=r_\mu e^{-i\phi_\mu},
		\qquad
		\Phi=\phi_A-\phi_B-\theta .
	\end{equation}
	The transverse amplitude dynamics then reads
	\begin{align}
		\dot r_A
		&=
		-\frac{\Gamma_A}{2}r_A
		+
		Vs_A^z r_B\sin\Phi ,
		\label{eq:SM_rA}
		\\
		\dot r_B
		&=
		-\frac{\Gamma_B}{2}r_B
		-
		Vs_B^z r_A\sin\Phi .
		\label{eq:SM_rB}
	\end{align}
	These equations show directly how the phase-dependent cross-feedback compensates the
	local transverse losses. For both amplitudes to remain finite, the same value of
	$\Phi$ must make the feedback term compensate damping in both ensembles. This is
	possible when $s_A^z$ and $s_B^z$ have opposite signs, consistent with the linear
	condition $\Delta_A\Delta_B<0$. Thus the locked relative phase is not an additional
	assumption; it is the phase selected by the feedback loop so that the transverse losses
	of the two ensembles are compensated simultaneously.
	
	Finally, the growth of the unstable transverse mode is saturated by the longitudinal
	dynamics. As the transverse amplitudes increase, the interaction term transfers
	polarization between the two ensembles and shifts $s_A^z$ and $s_B^z$ away from their
	stationary values. The system then reaches a nonlinear balance in which the collective
	anti-damping generated by cross-feedback is compensated by the local transverse damping.
	The resulting state is a stable oscillatory phase with finite transverse amplitudes and a
	fixed internal phase difference.
	
	The mechanism can therefore be summarized as follows. Local pump and decay prepare two
	stationary two-level ensembles with opposite longitudinal polarizations. Each ensemble
	alone has only damped transverse coherence. The inter-ensemble exchange interaction
	couples these two damped coherences into a cross-feedback loop. When the longitudinal
	polarizations are opposite, this loop generates collective anti-damping and destabilizes
	the stationary phase. Nonlinear longitudinal dynamics then saturates the growth, while
	the same feedback loop fixes the internal phase relation. This produces the interlocked
	oscillatory phase.

	\section{Cumulant hierarchy and two-time correlations}
	
	In this section we derive the correlation formalism used in the main text.
	The mean-field equations describe the dynamics of one-body collective variables
	and give a symmetry-broken oscillatory representative in the thermodynamic
	limit. For a finite system, however, the Liouvillian steady state remains
	symmetric, so the transverse order cannot be characterized by charged one-body
	expectation values alone, as in standard finite-size diagnostics of continuous
	time-crystalline order~\cite{Watanabe2015Absence,Iemini2018BoundaryTC,Tucker2018ShatteredTime}.
	We therefore formulate the order in terms of two-body correlations. We first
	derive the equal-time cumulant hierarchy and then use the same framework to
	obtain the two-time correlation functions through the quantum regression
	theorem.
	
	\subsection{Equal-time cumulant equations}
	
	To characterize correlations beyond mean field, we introduce the equal-time
	connected transverse correlations. For a general time-dependent state, these
	are defined as
	\begin{equation}
		C_{AA}^{+-}(t)
		=
		\langle \sigma_{A,i}^+(t)\sigma_{A,j}^-(t)\rangle
		-
		\langle \sigma_{A,i}^+(t)\rangle
		\langle \sigma_{A,j}^-(t)\rangle,
		\qquad i\neq j,
	\end{equation}
	\begin{equation}
		C_{BB}^{+-}(t)
		=
		\langle \sigma_{B,k}^+(t)\sigma_{B,l}^-(t)\rangle
		-
		\langle \sigma_{B,k}^+(t)\rangle
		\langle \sigma_{B,l}^-(t)\rangle,
		\qquad k\neq l,
	\end{equation}
	\begin{equation}
		C_{AB}^{+-}(t)
		=
		\langle \sigma_{A,i}^+(t)\sigma_{B,k}^-(t)\rangle
		-
		\langle \sigma_{A,i}^+(t)\rangle
		\langle \sigma_{B,k}^-(t)\rangle,
	\end{equation}
	\begin{equation}
		C_{AB}^{-+}(t)
		=
		\langle \sigma_{A,i}^-(t)\sigma_{B,k}^+(t)\rangle
		-
		\langle \sigma_{A,i}^-(t)\rangle
		\langle \sigma_{B,k}^+(t)\rangle .
	\end{equation}
	Permutation symmetry within each ensemble has been used, so the correlations
	do not depend on the specific particle indices, apart from the restrictions
	\(i\neq j\) and \(k\neq l\) for the intra-ensemble correlations.
	
	The connection between these correlations and the mean-field order parameter is
	fixed by the weak global \(U(1)\) symmetry of the Liouvillian, whose role in
	nonstationary dissipative phases is closely related to symmetry constraints on
	Liouvillian steady states~\cite{Buca2019Dissipation,Minganti2018SpectralTheory}. The symmetry is
	generated by
	\begin{equation}
		Q=S_A^z+S_B^z ,
	\end{equation}
	and the corresponding global phase rotation acts as
	\begin{equation}
		U_\phi=e^{-i\phi Q},
		\qquad
		\mathcal U_\phi[\rho]=U_\phi\rho U_\phi^\dagger .
	\end{equation}
	Here \(\phi\) is an arbitrary global rotation angle, and
	\(\mathcal U_\phi\) is the associated phase-rotation superoperator. The
	collective ladder operators transform as
	\begin{equation}
		U_\phi^\dagger S_\mu^+ U_\phi
		=
		e^{i\phi}S_\mu^+,
		\qquad
		U_\phi^\dagger S_\mu^- U_\phi
		=
		e^{-i\phi}S_\mu^- .
	\end{equation}
	The Liouvillian is weakly \(U(1)\)-symmetric in the sense that
	\begin{equation}
		\mathcal L[\mathcal U_\phi[\rho]]
		=
		\mathcal U_\phi[\mathcal L[\rho]]
	\end{equation}
	for arbitrary \(\rho\) and arbitrary \(\phi\). Therefore, if
	\begin{equation}
		\mathcal L(\rho_{\rm ss})=0,
	\end{equation}
	then
	\begin{equation}
		\mathcal L[\mathcal U_\phi[\rho_{\rm ss}]]=0 .
	\end{equation}
	For a finite system with a unique steady state, this implies
	\begin{equation}
		\mathcal U_\phi[\rho_{\rm ss}]
		=
		\rho_{\rm ss}.
	\end{equation}
	Consequently, any operator carrying nonzero \(U(1)\) charge has a vanishing
	expectation value in the finite-size steady state. In particular,
	\begin{equation}
		\langle S_A^+\rangle_{\rm ss}
		=
		\langle S_B^+\rangle_{\rm ss}
		=
		0 .
	\end{equation}
	The one-body transverse order is therefore invisible in the symmetric
	finite-size steady state.
	
	By contrast, the two-body products
	\begin{equation}
		S_\mu^+S_\nu^-,
		\qquad
		\sigma_{\mu,i}^+\sigma_{\nu,j}^-,
	\end{equation}
	carry zero net \(U(1)\) charge and are invariant under the same global phase
	rotation. We refer to such correlations as \(U(1)\)-neutral, or phase-neutral,
	correlations. They can remain finite in the symmetric steady state and
	therefore provide the appropriate finite-size diagnostics of transverse order.
	
	This can be seen by phase-averaging the mean-field oscillatory solution
	\begin{equation}
		s_\mu^+(t)
		=
		R_\mu e^{-i(\omega t+\phi_\mu)} .
	\end{equation}
	The phase average gives
	\begin{equation}
		\overline{s_\mu^+}=0,
	\end{equation}
	but leaves the neutral product finite:
	\begin{equation}
		\overline{s_\mu^+(s_\nu^+)^*}
		=
		R_\mu R_\nu e^{-i(\phi_\mu-\phi_\nu)} .
	\end{equation}
	Thus, in the finite-size steady state,
	\(C_{AA,{\rm ss}}^{+-}\) and \(C_{BB,{\rm ss}}^{+-}\) encode the
	phase-neutral transverse amplitudes of the two ensembles, while
	\(C_{AB,{\rm ss}}^{+-}\) carries the locked relative phase between them.
	
	The equations of motion for the two-body correlations generate three-body
	expectation values. We close the hierarchy by neglecting third-order connected
	cumulants, following the standard second-order cumulant-expansion strategy for
	open many-body systems~\cite{Kubo1962CumulantExpansion,Plankensteiner2022QuantumCumulants,Wang2025LocalDissipationBTC,Russo2025QDCTC}:
	\begin{equation}
		\langle ABC\rangle
		\simeq
		\langle AB\rangle\langle C\rangle
		+
		\langle AC\rangle\langle B\rangle
		+
		\langle BC\rangle\langle A\rangle
		-
		2\langle A\rangle\langle B\rangle\langle C\rangle .
	\end{equation}
	For the finite-size symmetric steady-state sector considered here, the charged
	one-body averages vanish. The connected transverse correlations therefore
	reduce to the corresponding two-body averages in this sector. Applying the
	above truncation and suppressing the time arguments of
	\(C_{\mu\nu}^{+-}(t)\) and \(s_\mu^z(t)\), we obtain the following reduced
	cumulant equations:
	\begin{align}
		\dot C_{AA}^{+-}
		&=
		-\Gamma_A C_{AA}^{+-}
		+iV s_A^z
		\left(
		e^{i\theta}C_{AB}^{+-}
		-
		e^{-i\theta}C_{AB}^{-+}
		\right),\label{eq:caa_eom}
		\\
		\dot C_{BB}^{+-}
		&=
		-\Gamma_B C_{BB}^{+-}
		+iV s_B^z
		\left(
		e^{-i\theta}C_{AB}^{-+}
		-
		e^{i\theta}C_{AB}^{+-}
		\right),\label{eq:cbb_eom}
		\\
		\dot C_{AB}^{+-}
		&=
		\left(
		i\Delta\omega
		-\frac{\Gamma_A+\Gamma_B}{2}
		\right)C_{AB}^{+-}
		\nonumber\\
		&\quad
		+iVe^{-i\theta}
		\left[
		\frac{s_B^z-s_A^z}{2N}
		-\frac{N-1}{N}C_{BB}^{+-}s_A^z
		+\frac{N-1}{N}C_{AA}^{+-}s_B^z
		\right],
		\\
		\dot C_{AB}^{-+}
		&=
		\left(
		-i\Delta\omega
		-\frac{\Gamma_A+\Gamma_B}{2}
		\right)C_{AB}^{-+}
		\nonumber\\
		&\quad
		-iVe^{i\theta}
		\left[
		\frac{s_B^z-s_A^z}{2N}
		-\frac{N-1}{N}C_{BB}^{+-}s_A^z
		+\frac{N-1}{N}C_{AA}^{+-}s_B^z
		\right].
	\end{align}
	Here
	\begin{equation}
		s_\mu^z(t)
		=
		\frac{\langle S_\mu^z(t)\rangle}{N},
		\qquad
		\Gamma_\mu=\gamma_\mu^++\gamma_\mu^-,
		\qquad
		\Delta\omega=\omega_A-\omega_B .
	\end{equation}
	
	Introducing
	\[
	\boldsymbol{\mathcal C}
	=
	\left(
	C_{AA}^{+-},
	C_{BB}^{+-},
	C_{AB}^{+-},
	C_{AB}^{-+}
	\right)^T ,
	\]
	the reduced cumulant equations can be written as
	\begin{equation}
		\dot{\boldsymbol{\mathcal C}}
		=
		M_c\boldsymbol{\mathcal C}+\mathbf b_c,
		\qquad
		M_c
		=
		\begin{pmatrix}
			M_{\rm intra} & M_{12}\\
			M_{21} & M_{\rm inter}
		\end{pmatrix}.
	\end{equation}
	The intra-ensemble block is
	\begin{equation}
		M_{\rm intra}
		=
		\begin{pmatrix}
			-\Gamma_A & 0\\
			0 & -\Gamma_B
		\end{pmatrix},
	\end{equation}
	and the inter-ensemble block is
	\begin{equation}
		M_{\rm inter}
		=
		\begin{pmatrix}
			i\Delta\omega-\frac{\Gamma_A+\Gamma_B}{2} & 0\\
			0 & -i\Delta\omega-\frac{\Gamma_A+\Gamma_B}{2}
		\end{pmatrix}.
	\end{equation}
	The coupling blocks read
	\begin{align}
		M_{12}
		&=
		iV
		\begin{pmatrix}
			s_A^z e^{i\theta} & -s_A^z e^{-i\theta}\\
			-s_B^z e^{i\theta} & s_B^z e^{-i\theta}
		\end{pmatrix},
		\\
		M_{21}
		&=
		iV\frac{N-1}{N}
		\begin{pmatrix}
			s_B^z e^{-i\theta} & -s_A^z e^{-i\theta}\\
			-s_B^z e^{i\theta} & s_A^z e^{i\theta}
		\end{pmatrix},
	\end{align}
	and the inhomogeneous term is
	\begin{equation}
		\mathbf b_c
		=
		\frac{iV}{2N}
		\begin{pmatrix}
			0\\
			0\\
			e^{-i\theta}(s_B^z-s_A^z)\\
			-e^{i\theta}(s_B^z-s_A^z)
		\end{pmatrix}.
	\end{equation}
	The asymmetric finite-\(N\) factors in \(M_{12}\) and \(M_{21}\) have a
	simple counting origin. The block \(M_{12}\) describes how inter-ensemble
	correlations feed back into the intra-ensemble correlations. In this case the
	interaction term contains a full sum over the opposite ensemble, and the
	resulting factor \(N\) cancels the Kac factor \(1/N\) in the Hamiltonian.
	Therefore no factor \((N-1)/N\) appears in \(M_{12}\). By contrast,
	\(M_{21}\) describes how intra-ensemble pair correlations feed into the
	inter-ensemble sector. Since \(C_{AA}^{+-}\) and \(C_{BB}^{+-}\) are defined
	for distinct particles, the particle already appearing in the cross correlator
	must be excluded from the same-ensemble sum. This leaves \(N-1\) terms and
	produces the prefactor \((N-1)/N\).
	
	The inhomogeneous term \(\mathbf b_c\) comes from the corresponding local
	self contributions that remain after this separation of collective pair terms.
	It is proportional to \(1/N\), and therefore represents a finite-size
	background source for \(U(1)\)-neutral inter-ensemble correlations. This term
	should not be identified as the origin of the thermodynamic OP, since it
	vanishes in the thermodynamic limit \(N\to\infty\). The finite phase-neutral
	correlations in the OP arise from the collective instability and correspond to
	the phase-averaged form of the symmetry-broken oscillatory state, as discussed
	above.
	\begin{figure}[t]
		\centering
		\includegraphics[width=0.75\linewidth]{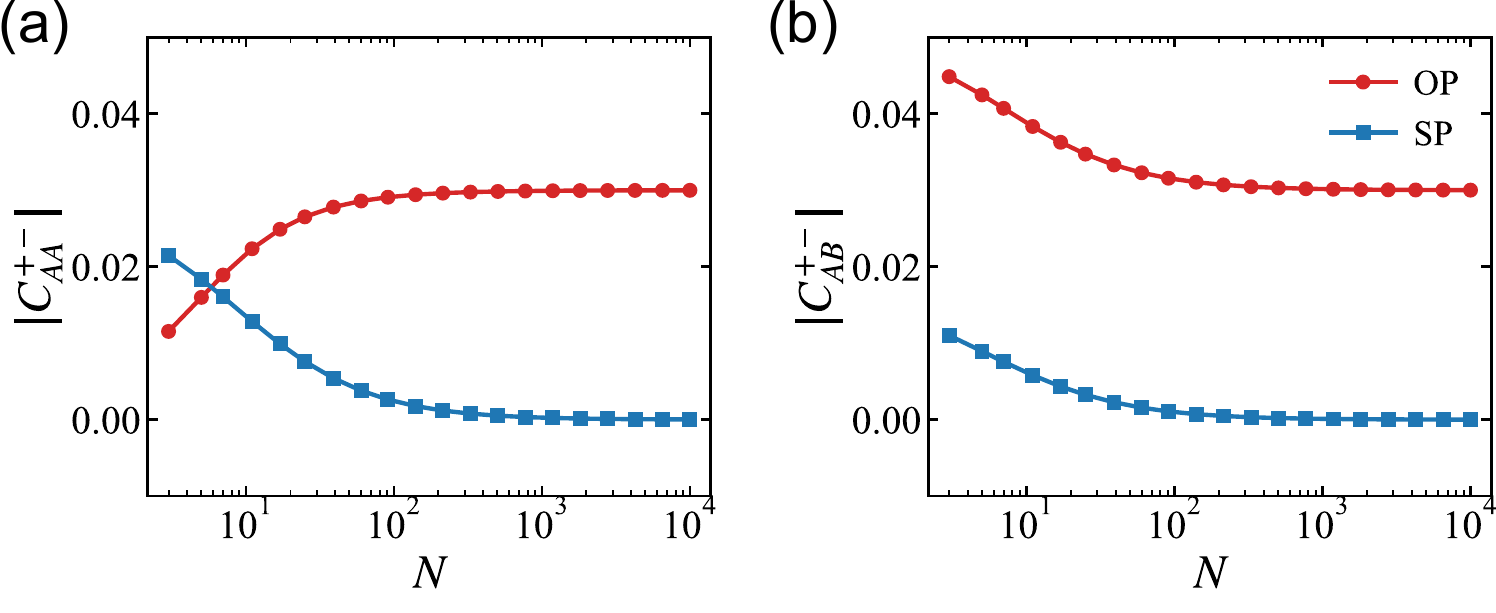}
		\caption{
			Steady-state phase-neutral equal-time correlations obtained from the reduced
			cumulant equations. 
			(a) Magnitude of the intra-ensemble transverse correlation
			\(|C_{AA,{\rm ss}}^{+-}|\). 
			(b) Magnitude of the inter-ensemble correlation
			\(|C_{AB,{\rm ss}}^{+-}|\). 
			In the stationary phase (SP), the normalized correlations decrease with
			increasing \(N\) and vanish in the thermodynamic limit. 
			In the oscillatory phase (OP), \(C_{AA,{\rm ss}}^{+-}\) and
			\(C_{AB,{\rm ss}}^{+-}\) approach finite values as \(N\to\infty\),
			representing the phase-neutral transverse amplitude and the locked
			inter-ensemble coherence, respectively.
		}
		\label{fig:SM_cumulant}
	\end{figure}
	
	This structure directly shows why the intra-ensemble correlations are not
	autonomous. If the inter-ensemble correlation sector is assumed to be absent,
	namely \(C_{AB}^{+-}=C_{AB}^{-+}=0\), Eqs.~\eqref{eq:caa_eom} and \eqref{eq:cbb_eom} reduce to
	\begin{equation}
		\dot C_{AA}^{+-}=-\Gamma_A C_{AA}^{+-},
		\qquad
		\dot C_{BB}^{+-}=-\Gamma_B C_{BB}^{+-}.
	\end{equation}
	Thus the intra-ensemble correlations decay under local dissipation alone.
	Finite thermodynamic intra-ensemble correlations can therefore be sustained
	only together with the inter-ensemble correlation sector. This provides the
	correlation-level signature of interlocking: temporal order inside each
	ensemble is dynamically tied to the inter-ensemble coherence rather than being
	formed autonomously.
	
	The steady-state equal-time correlations are obtained from
	\begin{equation}
		0
		=
		M_c^{\rm ss}\boldsymbol{\mathcal C}_{\rm ss}
		+
		\mathbf b_c^{\rm ss},
	\end{equation}
	where \(M_c^{\rm ss}\) and \(\mathbf b_c^{\rm ss}\) are obtained by replacing
	\(s_\mu^z\) with the steady-state value \(s_{\mu,\rm ss}^z\). This behavior is
	illustrated in Fig.~S1. In the SP, only finite-size correlations are present,
	and the normalized correlations vanish as \(N\to\infty\). In the OP, after
	the finite-size background becomes negligible, \(C_{AA,{\rm ss}}^{+-}\) and
	\(C_{AB,{\rm ss}}^{+-}\) approach finite values, representing the
	phase-neutral transverse amplitude and the locked inter-ensemble coherence,
	respectively.
	
	These steady-state correlations provide the initial conditions for the two-time functions discussed below.
	
	\subsection{Quantum regression theorem and two-time correlations}
	
	We define the normalized steady-state two-time correlation function as
	\begin{equation}
		G_{\mu\nu}(\tau)
		=
		\frac{1}{N^2}
		\lim_{t\to\infty}
		\left\langle
		S_\mu^+(t+\tau)S_\nu^-(t)
		\right\rangle .
	\end{equation}
	For a Markovian master equation, the quantum regression theorem (QRT)~\cite{Lax1963QuantumRegression,GardinerZoller2004QuantumNoise} gives
	\begin{equation}
		G_{\mu\nu}(\tau)
		=
		\frac{1}{N^2}
		{\rm Tr}
		\left[
		S_\mu^+
		e^{\mathcal L\tau}
		\left(
		S_\nu^- \rho_{\rm ss}
		\right)
		\right],
        \label{eq:SM_G_QRT}
	\end{equation}
	where \(\rho_{\rm ss}\) satisfies
	\begin{equation}
		\mathcal L(\rho_{\rm ss})=0 .
	\end{equation}
	This expression is used for the exact finite-size Liouvillian calculations of the two-time correlations in the main text.
	The normalization by \(N^2\) does not affect the decay rates or spectral peak positions, but makes the two-time correlations directly comparable to the equal-time correlations introduced above.
	
	The QRT expression in Eq.~\eqref{eq:SM_G_QRT} is exact for finite \(N\). Its long-time
	behavior can be understood from the Liouvillian spectrum, which provides a
	natural language for dissipative phase transitions~\cite{Minganti2018SpectralTheory}. Schematically, one
	may write
	\begin{equation}
		G_{\mu\nu}(\tau)
		=
		\sum_\alpha
		\mathcal W_{\mu\nu}^{(\alpha)}(N)
		e^{\lambda_\alpha(N)\tau},
	\end{equation}
	where \(\mathcal W_{\mu\nu}^{(\alpha)}(N)\) is the normalized spectral weight
	of the \(\alpha\)-th Liouvillian mode in the correlation function
	\(G_{\mu\nu}\). Consequently,
	\begin{equation}
		P_{\mu\nu}(\omega)
		=
		\left|
		\sum_\alpha
		\frac{
			\mathcal W_{\mu\nu}^{(\alpha)}(N)
		}{
			\lambda_\alpha(N)+i\omega
		}
		\right| .
	\end{equation}
	This expression makes explicit that the finite-size spectra depend on both the
	Liouvillian eigenvalues and the corresponding spectral weights. The eigenvalues
	\(\lambda_\alpha(N)\) determine the peak positions and linewidths, while
	\(\mathcal W_{\mu\nu}^{(\alpha)}(N)\) determines how strongly a given mode
	contributes to \(G_{\mu\nu}\). For a slow oscillatory pair
	\begin{equation}
		\lambda_{\rm slow}^{\pm}(N)
		=
		-\kappa(N)\pm i\Omega(N),
	\end{equation}
	the associated spectral peaks occur at frequencies set by \(\Omega(N)\), up to
	the sign convention in the Fourier transform, and have linewidth
	\begin{equation}
		\kappa(N)
		=
		-\mathrm{Re}\,\lambda_{\rm slow}^{\pm}(N).
	\end{equation}
	Thus the narrowing of the exact finite-size spectral peaks is controlled by
	the same Liouvillian decay rate discussed in the main text.
	
	To connect this exact QRT picture with the cumulant analysis, we now derive a
	reduced regression equation. Applying the same second-order truncation used for
	the equal-time correlations, and replacing the longitudinal operator by its
	steady-state value, gives
	\begin{equation}
		\frac{d}{d\tau}\boldsymbol{\mathcal G}_\nu(\tau)
		=
		M_G^{\rm ss}\boldsymbol{\mathcal G}_\nu(\tau), \qquad\nu=A,B
	\end{equation}
	where
	\begin{equation}
		\boldsymbol{\mathcal G}_\nu(\tau)
		=
		\begin{pmatrix}
			G_{A\nu}(\tau)\\
			G_{B\nu}(\tau)
		\end{pmatrix},
	\end{equation}
	and
	\begin{equation}
		M_G^{\rm ss}
		=
		\begin{pmatrix}
			i\omega_A-\Gamma_A/2
			&
			-iVe^{-i\theta}s_{A,\rm ss}^z
			\\
			-iVe^{i\theta}s_{B,\rm ss}^z
			&
			i\omega_B-\Gamma_B/2
		\end{pmatrix}.
	\end{equation}
	This matrix has the same transverse structure as the stability matrix
	\(M_*\), with the stationary fixed-point polarizations \(s_{\mu,*}^z\)
	replaced by the steady-state values \(s_{\mu,\rm ss}^z\).
	
	The absence of an additional explicit factor of \(N\) in \(M_G^{\rm ss}\)
	follows from the Kac scaling of the inter-ensemble interaction. The interaction
	contains the factor \(V/N\), while the collective sum in the coupled operator
	contributes a factor \(N\). For example, at the level of the reduced regression
	equation,
	\begin{equation}
		\frac{V}{N}S_A^z S_B^+
		\;\longrightarrow\;
		Vs_{A,\rm ss}^z S_B^+ ,
	\end{equation}
	and similarly for the \(B\) equation. Therefore, for \(N_A=N_B=N\), these
	factors cancel at leading order.
	
	The initial condition of the reduced regression equation is
	\begin{equation}
		\boldsymbol{\mathcal G}_\nu(0)
		=
		\frac{1}{N^2}
		\begin{pmatrix}
			\langle S_A^+S_\nu^-\rangle_{\rm ss}\\
			\langle S_B^+S_\nu^-\rangle_{\rm ss}
		\end{pmatrix}.
	\end{equation}
	For the off-diagonal components, this initial condition is directly determined
	by the steady-state inter-ensemble correlations. For example,
	\begin{equation}
		G_{AB}(0)=C_{AB,\rm ss}^{+-},
		\qquad
		G_{BA}(0)=C_{AB,\rm ss}^{-+},
	\end{equation}
	in the symmetric finite-size steady state where the charged one-body averages
	vanish. For the diagonal components, the collective correlator contains an
	additional local self term:
	\begin{equation}
		G_{AA}(0)
		=
		\left(1-\frac{1}{N}\right)C_{AA,\rm ss}^{+-}
		+
		\frac{1}{N}
		\langle \sigma_{A,i}^+\sigma_{A,i}^-\rangle_{\rm ss},
	\end{equation}
	and analogously for \(G_{BB}(0)\). Thus the finite-size dependence of the
	reduced spectra enters through the steady-state equal-time correlations
	contained in \(\mathbf G_\nu(0)\). In the SP, these normalized correlations
	scale as \(O(1/N)\), so the spectral weight vanishes in the thermodynamic
	limit. In the OP, they remain finite, yielding a finite spectral weight as
	\(N\to\infty\).
	
	Within the reduced regression equation, the frequency-resolved correlations are
	obtained from
	\begin{equation}
		\int_0^\infty d\tau\,
		e^{i\omega\tau}
		\boldsymbol{\mathcal G}_\nu(\tau)
		=
		-
		\left(
		M_G^{\rm ss}+i\omega I
		\right)^{-1}
		\boldsymbol{\mathcal G}_\nu(0),
	\end{equation}
	provided that the relevant eigenvalues have negative real parts for finite
	\(N\). To make the pole structure explicit in the reduced description, let
	\(\lambda_j^{\rm rdc}\) be the eigenvalues of \(M_G^{\rm ss}\). Then the
	reduced spectrum has the schematic form
	\begin{equation}
		P_{\mu\nu}^{\rm rdc}(\omega)
		=
		\left|
		\sum_j
		\frac{
			\mathcal W_{\mu\nu}^{(j),{\rm rdc}}
		}{
			\lambda_j^{\rm rdc}+i\omega
		}
		\right| ,
	\end{equation}
	where \(\mathcal W_{\mu\nu}^{(j),{\rm rdc}}\) is determined by the projection of
	the initial condition \(\boldsymbol{\mathcal G}_\nu(0)\) onto the \(j\)-th reduced regression
	mode. For a reduced slow mode
	\begin{equation}
		\lambda_s^{\rm rdc}
		=
		-\kappa_G+i\Omega_G,
	\end{equation}
	the corresponding reduced spectral peak is located at
	\begin{equation}
		\omega_{\rm peak}^{G}\simeq -\Omega_G,
	\end{equation}
	with linewidth
	\begin{equation}
		\kappa_G
		=
		-\mathrm{Re}\,\lambda_s^{\rm rdc}.
	\end{equation}
	Thus \(M_G^{\rm ss}\) identifies the collective spectral branch shared by
	\(G_{AA}\), \(G_{BB}\), \(G_{AB}\), and \(G_{BA}\), while
	\(\boldsymbol{\mathcal G}_\nu(0)\) determines the corresponding spectral weight in the
	reduced description. The exact finite-size linewidths and lifetime scaling,
	however, are controlled by the full Liouvillian eigenvalues in the QRT
	expression, in particular by \(\lambda_{\rm slow}(N)\).
	
	This should be distinguished from models with an exact strong dynamical symmetry, where finite-size one-body observables may oscillate persistently due to exact imaginary Liouvillian eigenvalues~\cite{Buca2019Dissipation}.
	Here the finite-size oscillatory modes have negative real parts, and persistent temporal order emerges through the closing of their decay rates in the thermodynamic limit.

	\section{Frequency locking and stochastic robustness}
	
	The correlation and regression formalism developed in Sec.~S2 provides
	a direct way to test how the interlocked OP responds to parameter
	imperfections.  In this section, we apply this framework to two types of
	perturbations.  We first vary the bare frequency detuning to determine
	the finite locking window of the collective spectral branch.  We then
	introduce stochastic fluctuations in the interaction strength and in the
	local pumping rates to test the dynamical stability of the locked
	oscillation.
	
	\subsection{Robustness against frequency detuning}
	
	We begin with a deterministic detuning scan.  The bare frequency
	mismatch is defined as
	\begin{equation}
		\Delta\omega=\omega_A-\omega_B .
	\end{equation}
	For each value of \(\Delta\omega\) and \(V\), we compute the diagonal
	two-time spectra of the two ensembles.  These spectra diagnose whether
	temporal order develops in ensemble \(A\), in ensemble \(B\), or only as
	a collective interlocked response of the coupled system.
	
	To characterize the temporal order of each ensemble separately, we
	evaluate the diagonal two-time spectra
	\begin{equation}
		P_{\mu\mu}(\omega)
		=
		\left|
		\int_0^\infty d\tau\,
		e^{i\omega\tau}
		\mathcal G_{\mu\mu}(\tau)
		\right|,
		\qquad
		\mu=A,B ,
	\end{equation}
	where \(\mathcal G_{\mu\mu}(\tau)\) is the normalized steady-state
	two-time correlation defined in Sec.~S2.  We define the corresponding
	peak spectral amplitude as
	\begin{equation}
		P_{\rm peak}^{\mu\mu}
		=
		\max_\omega P_{\mu\mu}(\omega).
	\end{equation}
	Because \(P_{AA}(\omega)\) and \(P_{BB}(\omega)\) probe the temporal
	order within ensembles \(A\) and \(B\), respectively, they provide a
	direct diagnostic of whether either ensemble can form an autonomous OP.
	
\begin{figure}[t]
		\centering
		\includegraphics[width=0.6\linewidth]{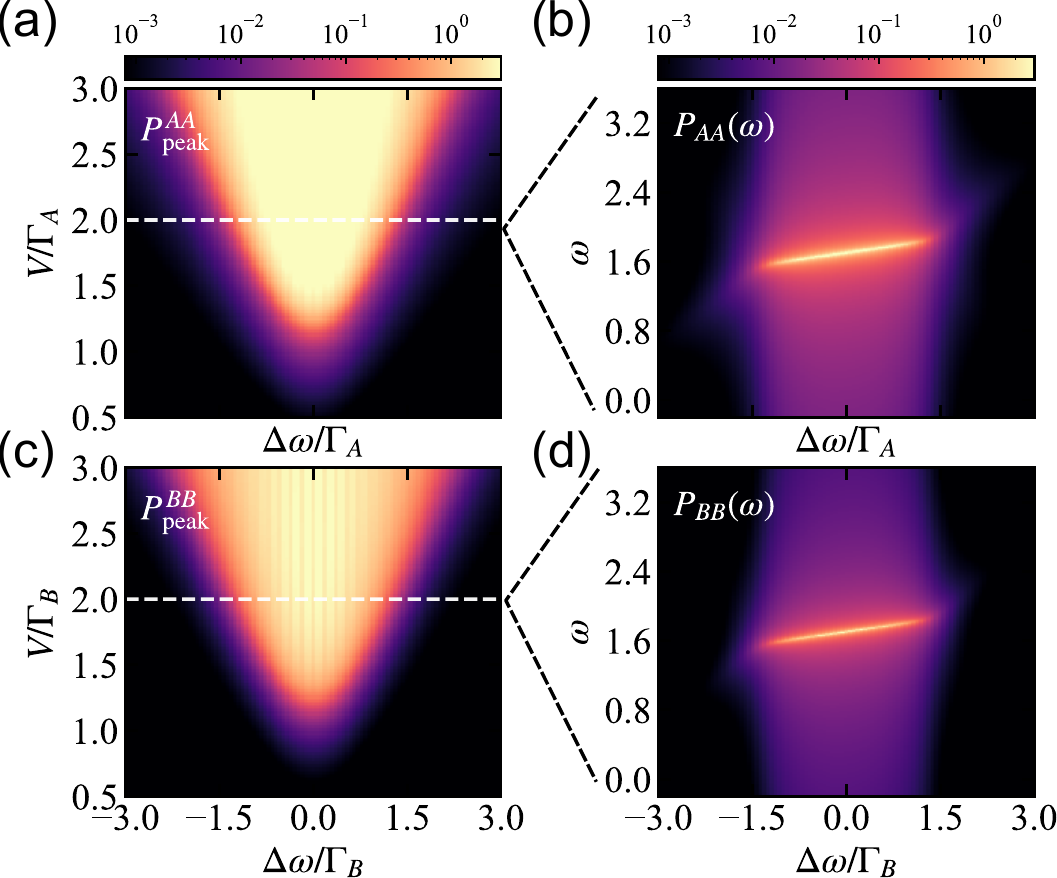}
		\caption{
			FIG.~S2. Robustness of the interlocked OP against frequency detuning
            for asymmetric damping rates \(\Gamma_A\neq\Gamma_B\).
            (a,c) Peak spectral amplitudes
            \(P^{AA}_{\rm peak}=\max_\omega P_{AA}(\omega)\) and
            \(P^{BB}_{\rm peak}=\max_\omega P_{BB}(\omega)\) as functions of the
            normalized detuning \(\Delta\omega/\Gamma_\mu\) and interaction
            strength \(V/\Gamma_\mu\), with
            \(\bar{\Gamma}=(\Gamma_A+\Gamma_B)/2\).  The dashed lines indicate the
            cut \(V/\Gamma_\mu=2\).  Both diagonal spectra are enhanced in the
            same tongue-shaped region, showing that the two ensembles enter the OP
            collectively.  (b,d) Frequency-resolved spectra \(P_{AA}(\omega)\) and
            \(P_{BB}(\omega)\) along the same cut.  Inside the locking window, both
            spectra select the same detuning-dependent collective branch, whose
            linear drift follows Eq.~\eqref{eq:SM_omega_c}.  Outside the locking window, no
            independent single-ensemble spectral branch develops.
		}
		\label{fig:SM_arnold}
	\end{figure}

	Figure~\ref{fig:SM_arnold}(a,c) shows \(P^{AA}_{\rm peak}\) and \(P^{BB}_{\rm peak}\) in the \((\Delta\omega/\Gamma_\mu,V/\Gamma_\mu)\) plane.  
    In this calculation we choose \(\Gamma_A\neq \Gamma_B\), so that the collective frequency is expected to vary with detuning according to Eq.~\eqref{eq:SM_omega_c}.
    Both spectra are enhanced inside the same tongue-shaped region centered around \(\Delta\omega=0\).  
    The width of this region grows with increasing interaction strength, showing that stronger inter-ensemble coupling stabilizes the collective oscillation against frequency mismatch.
    
    The same figure also clarifies the nature of the locking region.
    The enhanced regions of \(P_{\rm peak}^{AA}\) and \(P_{\rm peak}^{BB}\) coincide throughout the scanned parameter range.
    We do not observe a separate region in which only one of the two diagonal spectra develops a dominant collective peak.  
    Thus, outside the interlocked tongue, neither ensemble forms an independent OP at the level of its diagonal two-time spectrum.
    
    Figures~\ref{fig:SM_arnold}(b,d) show the full spectra
    \(P_{AA}(\omega)\) and \(P_{BB}(\omega)\) along the cut
    \(V/\Gamma_\mu=2\), indicated by the dashed lines in
    Figs.~\ref{fig:SM_arnold}(a,c).  Inside the locking window, both
    ensembles select the same detuning-dependent collective branch.  The
    ridge follows the analytical mean-field frequency
    \[
    \omega_c=\bar{\omega}
    -\frac{\Delta\omega}{2}
    \frac{\Gamma_A-\Gamma_B}{\Gamma_A+\Gamma_B},
    \]
    which becomes dispersive when \(\Gamma_A\neq\Gamma_B\).  Outside the
    locking window, the dominant collective peak is suppressed rather than
    splitting into two independent branches associated with the bare
    frequencies of the two ensembles.
	
	This behavior is qualitatively different from ordinary synchronization
	between two autonomous oscillators~\cite{Pikovsky2001Synchronization,Acebron2005KuramotoReview,Nadolny2023MacroscopicSync}.  In an ordinary synchronization
	scenario, each subsystem already possesses its own oscillatory branch
	before coupling locks their frequencies.  In the present system, by
	contrast, the diagonal spectra do not reveal independent single-ensemble
	OP regions outside the tongue.  The common spectral branch appears only
	inside the collective locking window.  Therefore the tongue in
	Fig.~\ref{fig:SM_arnold} should be interpreted as the detuning robustness
	region of a single interlocked OP, rather than as a synchronization
	region between two independently formed OPs.
	
	In the ideal model considered in this subsection, the detuning entering
	the locking diagram is the bare detuning \(\Delta\omega\).  In realistic
	implementations, residual intra-ensemble interactions can shift the
	effective detuning experienced by the collective mode.  This does not
	alter the intrinsic locking criterion, but changes how a given bare
	detuning maps into the locking window.  This effect is discussed in
	Sec.~S4.

	\subsection{Robustness against stochastic fluctuations}
	
	We next examine the stability of the interlocked OP against stochastic
	fluctuations.  We consider two representative noise sources:
	fluctuations of the inter-ensemble interaction strength and fluctuations
	of the local pumping rates.  The former perturbs the coupling that
	interlocks the two ensembles, while the latter perturbs the local
	gain-loss imbalance responsible for the collective instability.
	
	For interaction noise, we take
	\begin{equation}
		V(t)
		=
		V+\xi_V(t),
	\end{equation}
	where \(\xi_V(t)\) is a Gaussian white-noise process~\cite{GardinerZoller2004QuantumNoise} satisfying
	\begin{equation}
		\langle \xi_V(t)\rangle=0,
		\qquad
		\langle \xi_V(t)\xi_V(t')\rangle
		=
		2D_V\delta(t-t').
	\end{equation}
	
	For pumping noise, we consider independent fluctuations acting on the
	two ensembles,
	\begin{equation}
		\gamma_A^+(t)
		=
		\gamma_A^+
		+
		\xi_A(t),
		\qquad
		\gamma_B^+(t)
		=
		\gamma_B^+
		+
		\xi_B(t).
	\end{equation}
	Here \(\xi_A(t)\) and \(\xi_B(t)\) are independent Gaussian white-noise
	processes~\cite{GardinerZoller2004QuantumNoise} satisfying
	\begin{equation}
		\langle \xi_\mu(t)\rangle=0,
		\qquad
		\langle \xi_\mu(t)\xi_\nu(t')\rangle
		=
		2D_{\gamma_+}\delta_{\mu\nu}\delta(t-t'),
		\qquad
		\mu,\nu=A,B .
	\end{equation}
	Thus the pumping fluctuations of the two ensembles are uncorrelated,
	\begin{equation}
		\langle \xi_A(t)\xi_B(t')\rangle=0 .
	\end{equation}
	This choice represents local pumping noise rather than a common-mode
	fluctuation shared by the two ensembles.
	
	For each noise strength, the stochastic dynamics is simulated over
	multiple independent noise realizations generated using different random
	seeds.  The plotted values of \(Q_A\), \(Q_B\), and \(Q_{\rm phase}\)
	are averages over these realizations, while the error bars denote the
	corresponding standard deviations.  These error bars therefore quantify
	trajectory-to-trajectory fluctuations induced by the stochastic
	perturbations, rather than numerical integration uncertainties.  Small
	error bars indicate that the collective oscillation and phase locking
	are largely insensitive to the specific realization of the noise, whereas
	larger error bars signal enhanced realization dependence of the noisy
	dynamics.
	\begin{figure}[b]
		\centering
		\includegraphics[width=0.9\linewidth]{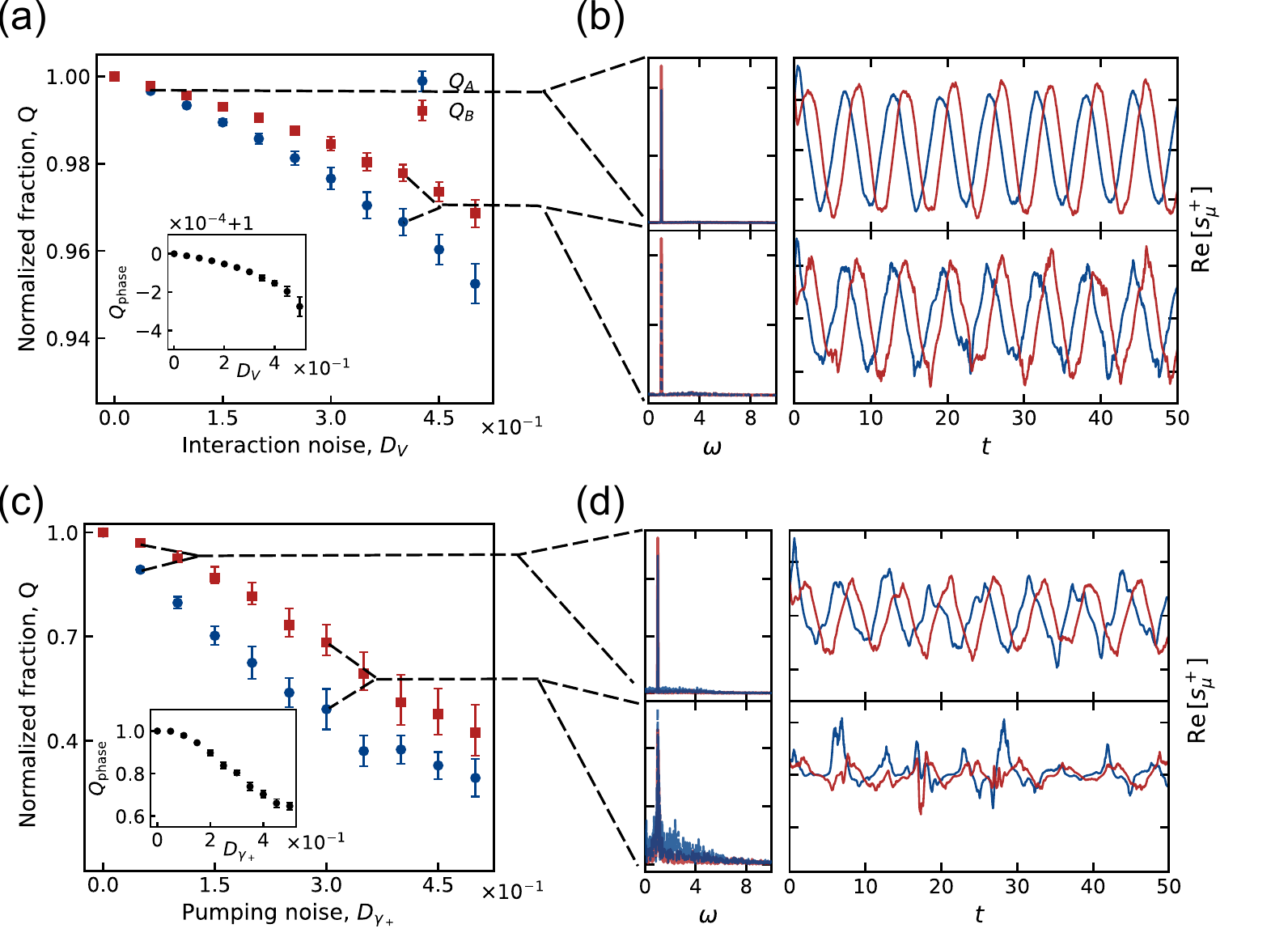}
		\caption{
			Robustness of the interlocked OP against stochastic fluctuations.
			(a) Noise-averaged normalized oscillation fractions \(Q_A\) and
			\(Q_B\) as functions of the interaction-noise strength \(D_V\).
			Error bars denote standard deviations over independent noise
			realizations generated with different random seeds.  The inset
			shows the phase-locking measure \(Q_{\rm phase}\).
			(b) Representative spectra and time-domain trajectories under
			interaction noise.
			The dominant spectral peak and locked oscillations remain stable.
			(c) Noise-averaged normalized oscillation fractions under
			independent pumping noise \(D_{\gamma_+}\).
			Error bars again denote standard deviations over independent noise
			realizations.  The inset shows the corresponding phase-locking
			measure.
			(d) Representative spectra and time-domain trajectories under
			independent pumping noise.
			Pumping noise gradually broadens the spectrum and suppresses the
			oscillation fraction, but a collective spectral peak remains
			visible for moderate noise strengths.
		}
		\label{fig:SM_noise}
	\end{figure}
	
	The response to interaction noise is shown in
	Fig.~\ref{fig:SM_noise}(a,b).  The normalized oscillation fractions
	\(Q_A\) and \(Q_B\) remain close to unity over the range of
	interaction-noise strengths considered.  The phase-locking measure
	\(Q_{\rm phase}\), shown in the inset, also remains close to unity.
	The small error bars indicate that different interaction-noise histories
	lead to nearly the same oscillation fraction and phase-locking quality.
	Consistently, the spectra retain a sharp dominant peak, and the
	time-domain trajectories continue to show regular locked oscillations.
	These results indicate that the interlocked OP is robust against
	moderate fluctuations of the inter-ensemble coupling.
	
	The response to independent pumping noise is shown in
	Fig.~\ref{fig:SM_noise}(c,d).  Compared with interaction noise,
	independent pumping noise has a stronger effect because it directly
	modifies the local gain-loss balance of each ensemble and also introduces
	relative fluctuations between the two ensembles.  As \(D_{\gamma_+}\)
	increases, the normalized oscillation fractions decrease and the phase
	coherence is gradually reduced.  The spectra broaden, and the
	time-domain trajectories become less regular.  The larger error bars at
	stronger pumping noise indicate that the dynamics becomes more
	realization dependent, consistent with intermittent degradation of the
	gain-loss balance and relative phase coherence.  Nevertheless, a
	dominant collective spectral peak remains visible over a finite range of
	pumping-noise strengths, showing that the interlocked OP survives
	moderate independent fluctuations of the local pumping processes.
	
	Together, Figs.~\ref{fig:SM_arnold} and \ref{fig:SM_noise} demonstrate
	the robustness of the interlocked time crystal in two complementary
	ways.  The diagonal spectra reveal a finite detuning window in which the
	two ensembles form a single collective OP, with no separate
	single-ensemble OP regions outside the tongue.  The stochastic
	simulations further show that, once formed, the interlocked OP remains
	stable against moderate fluctuations in both the interaction strength
	and the local pumping rates.

	\section{Intra-ensemble interactions and experimental detuning compensation}
	
	In Sec.~S3, we showed that the interlocked OP survives within a finite
	frequency-locking window in the plane of the detuning and the
	inter-ensemble coupling strength.  We now discuss how residual
	intra-ensemble interactions modify this picture in realistic
	implementations.
	
	The key point is that intra-ensemble interactions do not generate an
	independent oscillatory phase of either ensemble.  Instead, their leading
	effect is to renormalize the precession frequencies of the two ensembles.
	They should therefore be viewed as interaction-induced detuning shifts.
	As a result, the relevant quantity controlling entry into the locking
	window is not the bare detuning \(\Delta\omega\), but an effective
	detuning \(\Delta\omega_{\rm eff}\).  In this sense, residual
	intra-ensemble interactions can shift the locking window when it is
	expressed in terms of the bare detuning and may enlarge the
	experimentally accessible range of bare frequency mismatch.  The
	intrinsic locking criterion, however, remains the same when expressed in
	terms of \(\Delta\omega_{\rm eff}\).
	
	The model in the main text describes an idealized situation in which two
	dissipative ensembles are coupled by an inter-ensemble exchange
	interaction.  In this section, we denote the inter-ensemble coupling by
	\(V_{AB}\), corresponding to \(V\) in the main text, in order to
	distinguish it from residual intra-ensemble interactions \(V_A\) and
	\(V_B\).  Such intra-ensemble interactions are generally expected in
	realistic implementations.  For example, in cavity- or
	resonator-mediated platforms, the same electromagnetic mode that
	mediates the desired coupling between the two ensembles can also induce
	collective interactions among particles within each ensemble~\cite{Sorensen2002BadCavities,Norcia2018CavitySpinExchange,Li2022CollectiveSpinLight}.
	Therefore, the absence of intra-ensemble interactions should not be regarded
	as a necessary experimental requirement.
	
	To describe this situation, we consider the extended Hamiltonian
	\begin{equation}
		H
		=
		H_0
		+
		H_{AB}
		+
		H_{\rm intra},
	\end{equation}
	where
	\begin{equation}
		H_0
		=
		\omega_A S_A^z
		+
		\omega_B S_B^z ,
	\end{equation}
	and
	\begin{equation}
		H_{AB}
		=
		\frac{V_{AB}}{N}
		\left(
		e^{i\theta}S_A^+S_B^-
		+
		e^{-i\theta}S_A^-S_B^+
		\right).
	\end{equation}
	The residual intra-ensemble interactions are written as
	\begin{align}
		H_{\rm intra}
		&=
		\frac{V_A}{N}
		\sum_{i<j}
		\left(
		\sigma_{A,i}^+\sigma_{A,j}^-
		+
		\sigma_{A,i}^-\sigma_{A,j}^+
		\right)
		\nonumber\\
		&\quad
		+
		\frac{V_B}{N}
		\sum_{k<l}
		\left(
		\sigma_{B,k}^+\sigma_{B,l}^-
		+
		\sigma_{B,k}^-\sigma_{B,l}^+
		\right).
	\end{align}
	Here \(V_A\) and \(V_B\) characterize the residual collective exchange
	interactions inside ensembles \(A\) and \(B\), respectively.
	
	At the mean-field level, these intra-ensemble interactions modify the
	transverse equations as
	\begin{align}
		\dot{s}_A^+
		&=
		\left(
		i\omega_A
		-
		\frac{\Gamma_A}{2}
		-
		i\frac{N-1}{N}V_A s_A^z
		\right)s_A^+
		-
		iV_{AB}e^{-i\theta}s_A^z s_B^+,
		\\
		\dot{s}_B^+
		&=
		\left(
		i\omega_B
		-
		\frac{\Gamma_B}{2}
		-
		i\frac{N-1}{N}V_B s_B^z
		\right)s_B^+
		-
		iV_{AB}e^{i\theta}s_B^z s_A^+ .
	\end{align}
	The intra-ensemble terms have the same structure as the bare precession
	terms.  Therefore, to leading order, they renormalize the effective
	precession frequencies:
	\begin{align}
		\omega_A^{\rm eff}
		&=
		\omega_A
		-
		\frac{N-1}{N}V_A s_A^z,
		\\
		\omega_B^{\rm eff}
		&=
		\omega_B
		-
		\frac{N-1}{N}V_B s_B^z .
	\end{align}
	The effective detuning entering the collective instability is then
	\begin{equation}
		\Delta\omega_{\rm eff}
		=
		\omega_A^{\rm eff}
		-
		\omega_B^{\rm eff}.
	\end{equation}
	Using the main-text convention
	\begin{equation}
		\Delta\omega
		=
		\omega_A-\omega_B ,
	\end{equation}
	we obtain
	\begin{equation}
		\boxed{
			\Delta\omega_{\rm eff}
			=
			\Delta\omega
			-
			\frac{N-1}{N}
			\left(
			V_A s_A^z
			-
			V_B s_B^z
			\right)
		}.
	\end{equation}
	
	Near the transition from the stationary phase to the interlocked
	oscillatory phase, the transverse coherences vanish and the longitudinal
	polarizations approach their incoherent stationary values,
	\begin{equation}
		s_{A,{\rm sp}}^z
		=
		\frac{\Delta_A}{\Gamma_A},
		\qquad
		s_{B,{\rm sp}}^z
		=
		\frac{\Delta_B}{\Gamma_B}.
	\end{equation}
	Here
	\begin{equation}
		\Delta_\mu
		=
		\gamma_\mu^+
		-
		\gamma_\mu^-,
		\qquad
		\Gamma_\mu
		=
		\gamma_\mu^+
		+
		\gamma_\mu^- .
	\end{equation}
	Thus the linearized effective detuning is
	\begin{equation}
		\boxed{
			\Delta\omega_{\rm eff}^{\rm lin}
			=
			\Delta\omega
			-
			\frac{N-1}{N}
			\left(
			V_A\frac{\Delta_A}{\Gamma_A}
			-
			V_B\frac{\Delta_B}{\Gamma_B}
			\right)
		}.
	\end{equation}
	For the symmetric case \(V_A=V_B\equiv V_{\rm intra}\) and
	\(\Gamma_A=\Gamma_B\equiv\Gamma\), this becomes
	\begin{equation}
		\Delta\omega_{\rm eff}^{\rm lin}
		=
		\Delta\omega
		-
		\frac{N-1}{N}
		V_{\rm intra}
		\frac{\Delta_A-\Delta_B}{\Gamma}.
	\end{equation}
	
	The mean-field instability condition derived in Sec.~S1 remains valid
	after replacing the bare detuning \(\Delta\omega\) by the effective
	detuning \(\Delta\omega_{\rm eff}^{\rm lin}\).  The critical
	inter-ensemble coupling therefore becomes
	\begin{equation}
		V_{AB,c}^2
		=
		-
		\frac{(\Gamma_A\Gamma_B)^2}{\Delta_A\Delta_B}
		\left[
		\frac{1}{4}
		+
		\frac{
			(\Delta\omega_{\rm eff}^{\rm lin})^2
		}
		{
			(\Gamma_A+\Gamma_B)^2
		}
		\right].
	\end{equation}
	This expression has the same form as the critical coupling in the ideal
	model, with the replacement
	\(\Delta\omega\to\Delta\omega_{\rm eff}^{\rm lin}\).  Therefore residual
	intra-ensemble interactions do not introduce a new instability mechanism
	or a separate single-ensemble OP.  Their main role is to shift the
	effective detuning that enters the collective instability.
	
	For a fixed inter-ensemble coupling \(V_{AB}\), the locking condition can
	be written schematically as
	\begin{equation}
		|\Delta\omega_{\rm eff}^{\rm lin}|
		\lesssim
		\Delta\omega_{\rm lock}(V_{AB}),
	\end{equation}
	where \(\Delta\omega_{\rm lock}(V_{AB})\) denotes the locking range
	identified in Sec.~S3.  Equivalently, in terms of the bare detuning,
	\begin{equation}
		\left|
		\Delta\omega
		-
		\frac{N-1}{N}
		\left(
		V_A\frac{\Delta_A}{\Gamma_A}
		-
		V_B\frac{\Delta_B}{\Gamma_B}
		\right)
		\right|
		\lesssim
		\Delta\omega_{\rm lock}(V_{AB}) .
	\end{equation}
	Thus the intra-ensemble interactions shift the center of the accessible
	bare-detuning window.  Depending on the sign of this shift, they can
	either help bring a detuned system into the interlocked window or move it
	away from the optimal locking condition.
	
	This observation is useful for experimental design. In a realistic platform,
	one should treat the detuning as an effective quantity containing both the bare
	frequency mismatch and the interaction-induced shifts. If the residual
	intra-ensemble interactions shift the system away from the optimal condition,
	the bare detuning \(\Delta\omega\) can be tuned externally to compensate this
	shift.
	
	For example, if the two ensembles are subject to independently tunable magnetic
	fields, their transition frequencies may be written as
	\begin{equation}
		\omega_A(B_A)=\omega_A(0)+\eta_A B_A,
		\qquad
		\omega_B(B_B)=\omega_B(0)+\eta_B B_B .
	\end{equation}
	The bare detuning is then
	\begin{equation}
		\Delta\omega(B_A,B_B)
		=
		\Delta\omega(0)+\eta_A B_A-\eta_B B_B .
	\end{equation}
	Thus the relative detuning can be adjusted by local frequency control of the
	two ensembles. More generally, the same compensation can be achieved through
	differential Zeeman shifts, ac Stark shifts, microwave dressing, or local
	cavity detunings.
	
	The compensation condition can be written as
	\begin{equation}
		\Delta\omega(B_A,B_B)
		\simeq
		\frac{N-1}{N}
		\left(
		V_A\frac{\Delta_A}{\Gamma_A}
		-
		V_B\frac{\Delta_B}{\Gamma_B}
		\right),
	\end{equation}
	so that \(\Delta\omega_{\rm eff}^{\rm lin}\simeq0\). More generally, the
	requirement is not exact cancellation, but rather
	\begin{equation}
		\left|
		\Delta\omega_{\rm eff}^{\rm lin}
		\right|
		\lesssim
		\Delta\omega_{\rm lock}(V_{AB}),
	\end{equation}
	where \(\Delta\omega_{\rm lock}(V_{AB})\) denotes the finite locking range
	identified in Sec.~S3. Thus residual intra-ensemble interactions do not need to
	be eliminated. They can be incorporated into the effective detuning budget and
	compensated by external frequency control, provided that the compensated
	effective detuning remains inside the interlocked window.
	
	The range of this compensation mechanism is finite.  Since the
	interaction-induced shift is of order \(V_A\) or \(V_B\), residual
	intra-ensemble interactions cannot compensate arbitrarily large bare
	frequency mismatches.  Rather, they shift the bare-detuning interval that
	maps into the finite effective-detuning window of Sec.~S3.  The
	accessible range of bare detunings is therefore enlarged only up to the
	scale of the interaction-induced frequency shift,
	\begin{equation}
		|\Delta\omega|_{\rm comp}
		\sim
		\max(|V_A|,|V_B|),
	\end{equation}
	in addition to the intrinsic locking range set by \(V_{AB}\) and the
	dissipative rates.  Thus the experimental requirement is not the
	complete elimination of intra-ensemble interactions, but the ability to
	characterize their induced frequency shifts and compensate them by
	external frequency control.
	
	In cavity-mediated implementations, this viewpoint is particularly
	relevant~\cite{Sorensen2002BadCavities,Norcia2018CavitySpinExchange,Li2022CollectiveSpinLight}.  The same cavity mode that generates \(V_{AB}\) can also
	generate \(V_A\) and \(V_B\).  Instead of treating these terms as fatal
	imperfections, they should be included in the effective detuning budget.
	Experimentally, one can first calibrate the collective resonance
	frequencies of the two ensembles separately, estimate the
	intra-ensemble interaction-induced shifts, and then tune \(\Delta\omega\) using local or differential Zeeman shifts, Stark
	shifts, microwave dressing, or cavity detunings.  The interlocked time-crystal
	phase is expected to persist as long as the compensated effective
	detuning and residual inhomogeneity remain inside the collective locking
	window set by the available inter-ensemble coupling and dissipative
	rates.

\end{document}